\documentclass[traditabstract,longauth]{aa}
\usepackage{graphicx}
\usepackage{txfonts}
\begin{document}
   \title{The GALEX Ultraviolet Virgo Cluster Survey (GUViCS). }
   \subtitle{I: The UV luminosity function of the central 12 sq.deg.}
\titlerunning{The GALEX UV Virgo Cluster Survey (GUViCS). The UV luminosity function of the central 12 sq.deg.}
\authorrunning{Boselli et al.}
  \author{A. Boselli\inst{1}
          ,
	  S. Boissier\inst{1}
	  ,
	  S. Heinis\inst{1}
	  ,
	  L. Cortese\inst{2}
	  ,
	  O. Ilbert\inst{1}
	  ,
	  T. Hughes\inst{3} 
	  ,
	  O. Cucciati\inst{4}
	  ,
	  J. Davies\inst{3}
	  ,
	  L. Ferrarese\inst{5}
	  ,
	  R. Giovanelli\inst{6}
	  ,
	  M.P. Haynes\inst{6}
	  ,
	  M. Baes\inst{7}
	  ,
	  C. Balkowski\inst{8}
	  ,
	  N. Brosch\inst{9}
	  ,
	  S.C. Chapman\inst{10}
	  ,
	  V. Charmandaris\inst{11,12,13}
	  ,
	  M. S. Clemens\inst{14}
	  ,
	  A. Dariush\inst{3,15}
	  ,
	  I. De Looze\inst{7}
	  ,
	  S. di Serego Alighieri \inst{16}
	  ,
	  P.-A Duc\inst{17}
	  ,
	  P. R. Durrell\inst{18}
	  ,
	  E. Emsellem\inst{2,19}
	  ,
	  T. Erben\inst{20}
	  ,
	  J. Fritz\inst{7}
	  ,
	  D. A. Garcia-Appadoo\inst{21}
	  ,
	  G. Gavazzi\inst{22}
	  ,
	  M. Grossi\inst{23}
	  ,
	  A. Jord\'an \inst{24,25}
	  ,
	  K. M. Hess\inst{26}
	  ,
	  M. Huertas-Company\inst{8,27}
	  ,
	  L. K. Hunt \inst{16}
	  ,
	  B.R. Kent\inst{28}
	  ,
	  D. G. Lambas\inst{29}
	  ,
	  A. Lan\c{c}on\inst{30}
	  ,
	  L.A. MacArthur\inst{5,31}
	  ,
	  S.C. Madden\inst{17}
	  ,
	  L. Magrini \inst{16}
	  ,
	  S. Mei \inst{8,27}
	  ,
	  E. Momjian\inst{32}
	  ,
	  R. P. Olowin\inst{33}
	  ,
	  E. Papastergis\inst{6}
	  ,
	  M. W. L. Smith \inst{3}
	  ,
	  J.M. Solanes \inst{34}
	  ,
	  O. Spector\inst{9}
	  ,
	  K. Spekkens \inst{35}
	  ,
	  J. E. Taylor\inst{36}
	  ,
	  C. Valotto\inst{29}
	  ,
	  W. van Driel\inst{8}
	  ,
	  J. Verstappen\inst{7}
	  ,
	  C. Vlahakis\inst{37}
	  ,
	  B. Vollmer\inst{30}
	  ,
	  E.M. Xilouris\inst{38} 
        }

\institute{	
		Laboratoire d'Astrophysique de Marseille, UMR6110 CNRS, 38 rue F. Joliot-Curie, F-13388 Marseille France
             \email{Alessandro.Boselli@oamp.fr, Samuel.Boissier@oamp.fr, Sebastien.Heinis@oamp.fr, Olivier.Ilbert@oamp.fr}
         \and  
	 	European Southern Observatory, Karl-Schwarzschild Str. 2, 85748 Garching bei Muenchen, Germany
		\email{lcortese@eso.org}
	  \and
	 	School of Physics and Astronomy, Cardiff University, Queens Buildings The Parade, Cardiff CF24 3AA, UK
             \email{Jonathan.Davies@astro.cf.ac.uk, Thomas.Hughes@astro.cf.ac.uk, Ali.Dariush@astro.cf.ac.uk, Matthew.Smith@astro.cf.ac.uk}
	 \and
                INAF-Osservatorio Astronomico di Trieste, Via Tiepolo 11,  34143 Trieste,  Italy 
              \email{Cucciati@oats.inaf.it}
         \and
	 	Herzberg Institute of Astrophysics, National Research Council of Canada, 5071 West Saanich Road, Victoria, BC V8X 4M6, Canada 
	     \email{laura.ferrarese@nrc-cnrc.gc.ca, Lauren.MacArthur@nrc-cnrc.gc.ca}
	 \and
	 	Center for Radiophysics and Space Research, Space Sciences Bldg., Cornell University, Ithaca, NY 14853, USA
	     \email{riccardo@astro.cornell.edu, haynes@astro.cornell.edu, papastergis@astro.cornell.edu}
	 \and
	        Sterrenkundig Observatorium, Universiteit Gent, Krijgslaan 281 S9, B-9000 Gent, Belgium
	     \email{maarten.baes@ugent.be, ilse.delooze@ugent.be, jacopo.fritz@ugent.be, joris.verstappen@ugent.be}
	 \and
                GEPI, Observatoire de Paris, CNRS, Univ. Paris Diderot, 5 place Jules Janssen, 92195 Meudon, France
             \email{chantal.balkowski@obspm.fr, marc.huertas@obspm.fr, simona.mei@obspm.fr, wim.vandriel@obspm.fr}
	 \and
                The Wise Observatory and the Florence and Raymoond Sackler Department of Physics and Astronomy, Faculty of Exact Sciences, Tel Aviv University, Tel Aviv 69978, Israel
             \email{moah@wise.tau.ac.il, odedspec@wise.tau.ac.il}
	 \and
	        Institute of Astronomy, University of Cambridge, Madingley Road, Cambridge CB3 0HA, UK
             \email{schapman@ast.cam.ac.uk}
	 \and
	        University of Crete, Department of Physics and Institute of Theoretical \& Computational Physics, GR-71003 Heraklion, Greece
	 \and 
	        IESL/Foundation for Research \& Technology-Hellas, GR-71110 Heraklion, Greece
	 \and 
	        Chercheur Associ\'e, Observatoire de Paris, F-75014 Paris, France
	     \email{vassilis@physics.uoc.gr}  
	 \and
	        INAF-Osservatorio Astronomico di Padova, Vicolo dell'Osservatorio 5, 35122 Padova, Italy
             \email{marcel.clemens@oapd.inaf.it}
	 \and
	        School of Astronomy, Institute for Research in Fundamental Sciences (IPM), PO Box 19395-5746, Tehran, Iran
	 \and
	        INAF-Osservatorio Astrofisico di Arcetri, Largo E. Fermi, 5, 50125, Firenze, Italy\\
             \email{sperello@arcetri.astro.it, hunt@arcetri.astro.it, laura@arcetri.astro.it}
	 \and
	        Laboratoire AIM Paris-Saclay, CNRS/INSU CEA/Irfu Universit\'e Paris Diderot, 91191 Gif sur Yvette cedex, France
             \email{paduc@cea.fr, smadden@cea.fr}
	 \and
	        Department of Physics and Astronomy, Youngstown State University, Youngstown, OH 44512 USA
             \email{prdurrell@ysu.edu}
	 \and
	        Universit\'e Lyon 1, Observatoire de Lyon, Centre de Recherche Astrophysique de Lyon and Ecole Normale Sup\'erieure de Lyon, 9 avenue Charles Andr\'e, F-69230 Saint-Genis Laval, France
             \email{emsellem@obs-unic-lyon1.fr}
	 \and
	        Argelander-Institut f\"ur Astronomie, Auf dem H\"ugel 71, D-53121 Bonn, Germany
             \email{terben@astro.uni-bonn.de}
	 \and
	        ESO, Alonso de Cordova 3107, Vitacura, Santiago, Chile
             \email{dgarcia@eso.org}
	 \and
	        Universita' di Milano-Bicocca, piazza della Scienza 3, 20100, Milano, Italy
             \email{giuseppe.gavazzi@mib.infn.it}
	 \and
	        CAAUL, Observat\'orio Astron\'omico de Lisboa, Universidade de Lisboa, Tapada de Ajuda, 1349-018, Lisboa, Portugal          
             \email{grossi@oal.ul.pt}
	 \and
	        Departamento de Astronom\'ia y Astrof\'isica, Pontificia Universidad Cat\'olica de Chile, 7820436 Macul, Santiago, Chile 
	     \email{ajordan@astro.puc.cl} 
	 \and
		Harvard-Smithsonian Center for Astrophysics, 60 Garden Street, Cambridge, MA 02138, USA 
	 \and
	 	Department of Astronomy, University of Wisconsin-Madison, 475 N. Charter St, Madison WI 53706, USA
             \email{hess@astro.wisc.edu}
         \and
                Universit\'e Paris Diderot, 75205 Paris Cedex 13, France
	 \and
	 	Jansky Fellow of the National Radio Astronomy Observatory, 520 Edgemont Road, Charlottesville, VA 22901
             \email{bkent@nrao.edu}
	 \and
	 	Instituto de Astronom\'{\i}a Te\'orica y Experimental, UNC-CONICET, Cordoba, Argentina
             \email{dgl@mail.oac.uncor.edu, val@mail.oac.uncor.edu}
	 \and
	 	Observatoire astronomique de Strasbourg, Université de Strasbourg, CNRS, UMR 7550, 11 rue de l'Université, F-67000 Strasbourg, France
             \email{ariane.lancon@astro.unistra.fr, Bernd.Vollmer@astro.unistra.fr}
	 \and
	 	Department of Physics \& Astronomy, University of Victoria, Victoria, BC V8P 1A1, Canada
	 \and
	 	National Radio Astronomy Observatory, P. O. Box O, Socorro, NM 87801, USA
             \email{emomjian@nrao.edu}
	 \and
	 	Saint Mary's College of California, Department of Physics and Astronomy, Moraga, CA, USA 94575,
             \email{rpolowin@stmarys-ca.edu}
	 \and
	 	Departament d'Astronomia i Meteorologia and Institut de Ci\`encies del Cosmos, Universitat de Barcelona, Mart\'{\i} i Franqu\`es 1, E-08028 Barcelona, Spain
             \email{jm.solanes@ub.edu}
	 \and
	 	Royal Military College of Canada, P.O. Box 17000, Station Forces, Kingston, Ontario, K7K 7B4 Canada
	     \email{Kristine.Spekkens@rmc.ca}
	 \and
	        Department of Physics and Astronomy, 200 University Avenue West, Waterloo, Ontario, Canada N2L1W7
             \email{taylor@uwaterloo.ca}
	 \and
	   	Departamento de Astronomia, Universidad de Chile, Casilla 36-D, Santiago, Chile
             \email{vlahakis@das.uchile.cl}
	 \and
	 	Institute of Astronomy and Astrophysics, National Observatory of Athens, I. Metaxa and Vas. Pavlou, P. Penteli, GR-15236 Athens, Greece 
            \email{xilouris@astro.noa.gr}	    
}

   \date{}

 
  \abstract  
{The GALEX Ultraviolet Virgo Cluster Survey (GUViCS) is a complete blind survey of the Virgo cluster covering $\sim$ 40 sq. deg. 
in the far UV (FUV, $\rm\lambda_{eff}=1539\AA, \Delta \lambda=442\AA$) and $\sim$ 120 sq. deg. in the near UV (NUV, $\rm
\lambda_{eff}=2316\AA, \Delta \lambda=1060\AA$). The goal of the survey is to study the
ultraviolet (UV) properties of galaxies 
in a rich cluster environment, spanning a wide luminosity range from giants to dwarfs, and regardless of prior knowledge of their star
formation activity. The UV data will be combined with those in other bands (optical: NGVS; far-infrared - submm: HeViCS; HI: ALFALFA) and with 
our multizone chemo-spectrophotometric models of galaxy evolution to make a complete 
and exhaustive study of the effects of the environment on the evolution of galaxies in high density regions.
We present here the scientific objectives of the survey, describing the observing strategy and briefly discussing different
data reduction techniques. Using UV data already in-hand for the central 12 sq. deg. we determine the FUV and NUV 
luminosity functions of the Virgo cluster core for all cluster members and separately for early- and late-type galaxies and compare it
to the one obtained in the field and other nearby clusters (Coma, A1367). This analysis shows that the FUV and NUV luminosity
functions of the core of the Virgo clusters are flatter ($\alpha$ $\sim$ -1.1) than those determined in Coma and A1367.
We discuss the possible origin of this difference.
 }
   {}
   {}
   {}
   {}
   {}


   \maketitle
%

\section{Introduction}

Observations of galaxies at different redshift have shown that galaxy mass is the principal driver of their evolution 
(Gavazzi et al. 1996; Cowie et al. 1996; Boselli et al. 2001).
The existence of systematic differences between galaxies inhabiting high-(clusters)
and low density (field) regions, among which the most evident is certainly the morphology segregation effect (Dressler 1980),
indicates, however, that the environment must have played a significant role in shaping galaxy evolution 
(Boselli \& Gavazzi 2006; Poggianti et al. 2009).
It is still unclear whether the seeds of galaxy formation in high density environments are
different from those in low density regions, therefore leading to different galaxy populations,  
or whether cluster and field galaxies are originally similar objects that are subsequently
modified by the harsh cluster environment (``Nurture or Nature''? Gavazzi et al. 2010). 
A critical step in furthering our understanding of these issues, is to compare the predictions of cosmological models 
and theories of structure formation and galaxy evolution
with multi-frequency observations of galaxies covering their entire ``fossil'' sequence, 
from their epoch of formation to fully evolved local systems.\\     
Studying the local universe is of paramount importance since it represents the end point of galaxy evolution.  
In the local universe most of the major baryonic components of galaxies (gas, stars, dust)
are accessible to modern instrumentation and the contribution of different galaxy components (nuclei, bulges, discs, spiral arms...)
can be separated. 
In this perspective the local universe is also ideal for gathering information 
on dwarf galaxies, the most common in the universe: because of their fragility due to a shallow potential
well, dwarfs are expected to be easily perturbed (Moore et al. 1998; Mastropietro et al. 2005;
Boselli et al. 2008a) and are therefore ideally suited to study the effects of environment on the evolution of galaxies.\\
Ironically, a global and complete view of high density environments in the local universe
is still lacking since clusters of galaxies, with their linear dimensions of a few Mpc,
have angular sizes that can reach up to $\sim$ 100 sq. deg. at a distance of $\sim$ 20 Mpc (Virgo, Fornax),
and are thus difficult to cover at all frequencies, even with modern instrumentation. They
indeed require panoramic detectors with a very large fields of view such as MegaPrime on CFHT (1 sq. deg.) 
and GALEX (circular field of 1 deg. diameter; 1 deg. at the distance of Virgo corresponds to $\sim$ 290 kpc).\\
This contribution will be focused on the Virgo cluster, the dominant mass concentration in the local universe and the largest
collection of galaxies within 35 Mpc. With the aim of studying the role of the environment
on the evolution of galaxies, several blind surveys are
covering this region at an unprecedented depth.
There are several reasons why Virgo has been chosen for these studies: it is a close, rich cluster, 
whose distance (16.5 Mpc; Mei et al. 2007) is such that galaxies spanning a wide range in morphology and luminosity can be studied, 
from giant spirals and ellipticals down to dwarf irregulars, 
blue compact dwarfs (BCDs) and dwarf ellipticals (dE) and spheroidals (dS0).
Furthermore, Virgo is still in the process of being assembled, so that a wide range of processes 
(ram-pressure stripping, tidal interactions, harassment 
and pre-processing) are still taking place. 
Three surveys are of particular relevance for these studies: the Next Generation Virgo Cluster Survey (NGVS, Ferrarese et al. 2011), 
The Herschel Virgo Cluster Survey (HeViCS, Davies et al. 2010), and the Arecibo Legacy Fast ALFA survey (ALFALFA, Giovanelli et al. 2005). \\
The Next Generation Virgo Survey (NGVS)\footnote{https://www.astrosci.ca/NGVS/\\
The\_Next\_Generation\_Virgo\_Cluster\_Survey/Home.html}
(Ferrarese et al., in preparation) is an optical ($ugriz$) survey covering 104 deg$^2$ of the Virgo 
cluster with MegaPrime on the CFHT to a point-source depth of $g$ $\sim$ 25.7 mag and
a corresponding surface brightness of $\mu_{g}$ $\sim$ 29 mag arcsec$^{-2}$. 
The survey, which is now starting its third year of operations, will be completed in 2012. 
The goals of the NGVS are the study of faint end slope of the galaxy luminosity function, 
the characterization of galaxy scaling relations over a dynamic range of 7 orders of magnitude in mass, and the study of the 
diffuse and discrete intracluster population.\\
The Herschel Virgo Cluster Survey (HeViCS)\footnote{http://www.hevics.org/} (Davies et al. 2010)
is a blind far-IR survey of 60 deg$^2$ in five photometric bands from 100 to 500 $\mu$m with PACS and SPIRE on 
the Herschel Space Observatory down to the confusion limit (at 250 $\mu$m; 286 hrs allocated as an open time key program on Herschel). The goal of this survey is to study the dust properties of 
cluster galaxies, including the extended dust distributed around galactic discs or associated with tidal debris
and low surface brightness galaxies, and to reconstruct the far-IR luminosity
function as well as to detect dust in the intra cluster medium.\\
   \begin{figure}
   \centering
   \includegraphics[width=9.5cm]{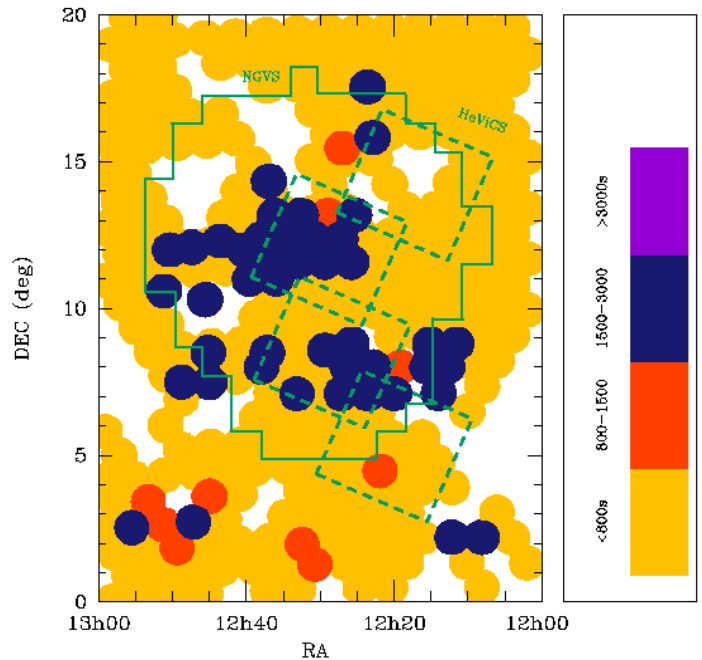}
   \caption{The Virgo cluster region as covered by GALEX in the FUV band. Circles of 1 deg. diameter indicate each single pointing, with 
   colours coded according to the exposure time.
   The region covered by NGVS is indicated by a green solid line while that mapped by HeViCS by a green dashed line.
   The entire 0$^o$ $<$dec$<$ 18$^o$ range has been covered by ALFALFA.}
   \label{guvics1}%
   \end{figure}
The Arecibo Legacy Fast ALFA survey (ALFALFA)\footnote{http://egg.astro.cornell.edu/alfalfa/}
is a blind HI survey covering 7000 deg$^2$ in the declination range 0$^o$$<$ dec$<$32$^o$ and velocity range 
-1600$<$ $vel$ $<$ 18000 km s$^{-1}$ with a spectral resolution of 5 km s$^{-1}$ down to a sensitivity limit of 2.4 mJy,
corresponding to $\sim$ 10$^{7.5}$ M$\odot$ at the distance of Virgo (Giovanelli et al. 2005).
The survey, which has already completed the Virgo cluster region,
has been designed to provide the basis for studies of the dynamics of galaxies within 
nearby superclusters, allow measurement of the HI diameter and mass function, and enable a first wide-area blind 
search for local HI tidal features and HI absorbers.\\
These surveys contain a wealth of information on the stellar population of Virgo galaxies, 
and on the properties of the neutral gas (available for star formation) and dust (produced during stellar evolution) in the cluster. 
A major ingredient, however, is still lacking for a complete study 
of the evolution of galaxies in clusters:
the present day star formation activity. This can be determined from the UV flux emitted by the youngest stellar
population, provided that dust extinction can be accurately determined, for instance, using the far-IR to UV flux ratio (Cortese et al. 2008a) \footnote{The
FAUST instrument (Deharveng et al. 1991) only provided a shallow FUV survey of the Virgo cluster with the detection of the brightest $\sim$ 40 objects.}. 
Furthermore, our previous studies have proved that combining multifrequency spectrophotometric data 
covering the entire electromagnetic spectrum, from the far-UV to radio
wavelengths, with multizone models of galaxy evolution, is one of the most reliable and powerful ways to reconstruct 
the evolution of galaxies in clusters. The galaxy-cluster interaction
responsible for the gas removal in the outer discs of spirals truncates the star formation activity producing 
reversed color gradients with respect to normal, unperturbed objects (Boselli et al. 2006). 
In dwarf, star forming systems, models indicate that the gas removal is so efficient in
stopping the activity on short time scales ($<$ 2 Gyr), producing objects with
structural and spectrophotometric properties similar to those of dwarf ellipticals (Boselli et al. 2008a,b).
Furthermore, the UV radiation is the most important heating source of the interstellar medium, and its quantification 
is critical for constraining the physical equilibrium between the different components of the ISM. Contrary to star forming systems,
in massive ellipticals the UV emission is dominated by the old stellar population (UV upturn; O'Connell 1999; Deharveng et al. 2002; Boselli et al. 2005a; Donas et al. 2007)
although some residual star formation can still be present.\\
Far-ultraviolet (FUV) and near-ultraviolet (NUV) GALEX observations of the Virgo area are thus mandatory for the success of this 
ambitious project of a complete and coherent study of the Virgo cluster, a milestone towards the understanding of
the relative role of mass and environment on the evolution of galaxies.
The goal of this paper is to present the GALEX Ultraviolet Virgo Cluster Survey (GUViCS)\footnote{http://galex.oamp.fr/guvics/},
designed specifically to cover the entire Virgo cluster region in the ultraviolet (UV) spectral domain. We present here all technical aspects
concerning the selected fields and the observing strategy, with a brief description of the data reduction procedures, with particular attention to the
standard GALEX pipeline tuned for the dominating background sources but limited for extended sources. The detailed description of
the flux extraction technique of Virgo cluster members will be presented in a forthcoming communication.
Using the data of the central 12 sq. deg. we construct the FUV and NUV luminosity functions of the core of the Virgo cluster.

\section{The GUViCS survey}

\begin{figure}
   \centering
   \includegraphics[width=9.5cm]{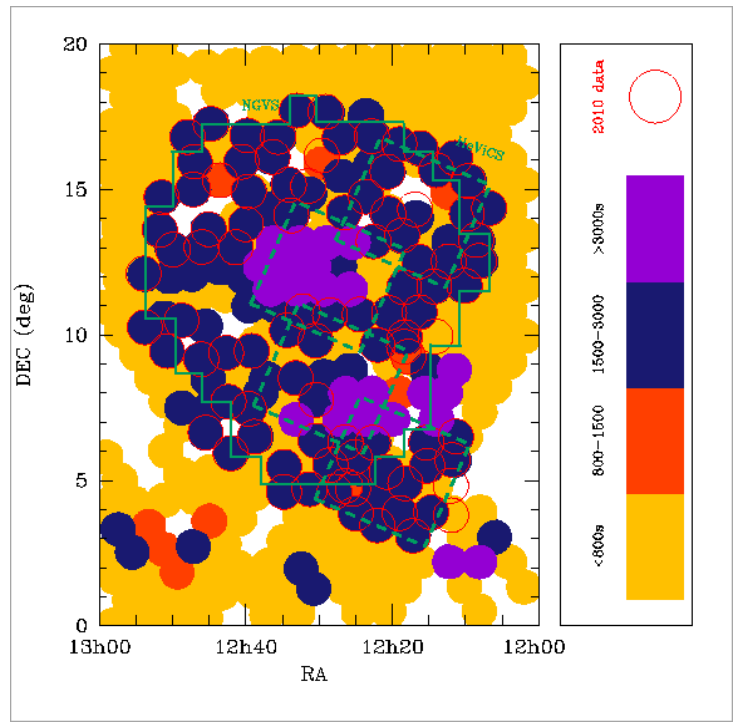}
   \caption{Same as Fig. \ref{guvics1} but for the NUV band. The GUViCS pointings are marked with a red circle.}
   \label{guvics2}%
   \end{figure}

Previous observations of nearby galaxies have shown that the low surface brightness features associated with low HI column densities, 
such as low surface brightness galaxies (Boissier et al. 2008), dwarf ellipticals
(Boselli et al. 2005a; Boselli et al. 2008a), 
the outskirts of spiral galaxies (Thilker et al. 2005) or 
tidal debris (Boselli et al. 2005b; Duc et al. 2007; Boquien et al. 2007, 2009), have UV surface brightnesses in both the FUV and NUV bands of $\sim$ 27.5-28 mag arcsec$^{-2}$, 
thus undetectable in the GALEX All-Sky Imaging Survey, which 
is limited to $\sim$ 26 mag arcsec$^{-2}$. 
Medium Deep Imaging Survey (MIS; 1 orbit/pointing, $\sim$ 1500 sec) are thus necessary to 
study the effects of the environment on cluster galaxies. 
MIS observations of the Virgo cluster have shown that while almost all star-forming galaxies catalogued in the VCC (Binggeli et al. 1985) down 
to $m_B$ $\sim$ 19 can be seen with 1 orbit observations, 
only the brightest dwarf ellipticals ($m_B$ $\sim$ 16.5) can be detected. The GUViCS project has been designed to 
complete the UV GALEX coverage of the Virgo cluster at a depth similar to the GALEX MIS. 
Excluding a few positions affected by the presence of bright stars saturating the detector,
94 new pointings have been selected in order to cover $\sim$ 90 \% of the area observed by the NGVS or the HeViCS surveys. This
results in a $\sim$ 120 square degree coverage centered on M87 (see Figs. \ref{guvics1} and \ref{guvics2}). 
In this area, the mean number of NUV and FUV detections per field
at the depth of the MIS is $\sim$ 10000 and 1500 respectively, with the majority being background galaxies (see sect. 5).
Considering the number of cluster objects catalogued in the VCC within the selected area,
we estimate that due to these new observations the numbers of UV detected Virgo members 
can be increased to $\sim$ 450 (late-type) and $\sim$ 260 (early-type) respectively (43 and 92 have been already detected in the central 12 deg$^2$
analyzed in this work). This number will certainly increase 
once the data are cross-correlated with the position of new galaxies discovered by NGVS, allowing the detection of peculiar objects such as
compact galaxies, unresolved in UV (GALEX has a resolution of 4-5 arcsec) but resolved by NGVS, and very low surface brightness systems undetected by optical surveys, 
or star forming tidal debris such as the UV tidal tail of NGC 4438 (Boselli et al. 2005b). This significant increase in the statistics of cluster members
is mandatory to extend previous analyses (Boselli et al. 2005a,b, 2006, 2008a,b; Lisker \& Han 2008; Kim et al. 2010), which were limited 
in the central 12 deg$^2$, to a narrow strip in the southern region of the 
cluster, or to some pointed observations on a few selected objects. \\
The availability of contiguous fields will allow us to have
a complete view of the cluster, from the densest core
to its outskirts. Furthermore, we can study the properties of galaxies within the various cluster dynamical substructures, from
Virgo A, the virialized region within the cluster, 
dominated by quiescent systems or gas deficient spirals, to Virgo B, the M, W and W' clouds and the Southern extension, these last
regions dominated by freshly infalling spirals undergoing their first interaction with the cluster environment (Gavazzi et al. 1999). These substructures,
undersampled in the previous GALEX surveys of Virgo, are of primary importance since they are the analogues of the groups that
merged in the past to form rich clusters, where pre-processing is probably active. This is still a poorly known process 
often proposed as the main driver of past galaxy 
evolution in high density regions. The significant increase in statistics will also allow us to make detailed analysis
of different sub-categories of quiescent (dwarf ellipticals and spheroidals, nucleated and not) and star-forming objects
(Im, BCD, pec, amorphous..) the origin of which is uncertain (Lisker et al. 2008; Cote et al. 2006).
The Balmer break falls in the $u$ band for $z<$0.25, therefore, GALEX data are 
crucial to sample the spectra of galaxies shortward of the break, thus helping to separate Virgo galaxies from background objects 
and removing any degeneracy with the morphological type. Combining UV data with those obtained in the optical bands by NGVS we will improve the photo-$z$ accuracy at 
$z<$0.5 by a factor of 1.5-2 and limit the fraction of $z<$0.25 galaxies with a photo-$z$ wrongly estimated at $z>$
0.4 (see Fig.3 of Niemack et al. 2009).\\
UV data will be used to determine, for the first time, the UV luminosity function of a nearby cluster down to $M_{UV}$=-10, a limit  $\sim$ 4
mag deeper than previously conducted (Cortese et al. 2005, 2008b), allowing us to sample both the quiescent and star forming galaxy population 
down to dwarfs. UV data will then be used to reconstruct the 
extinction corrected UV to sub-mm SED of Virgo cluster galaxies that, with the help of our models, will 
constrain the evolution of galaxies in the cluster for both resolved and unresolved objects (Boselli et al. 1998; 2003; 2010).
The UV color of early type galaxies will be analyzed to determine whether their emission is due to recent star formation activity or is
rather related to an old stellar population (UV upturn; Boselli et al. 2005a). 
We will study the UV-optical color magnitude relation, that combined with our spectrophotometric 
models of galaxy  evolution, will be used to understand the origin of the red sequence in clusters 
(Boselli et al. 2008a; Hughes \& Cortese 2009; Cortese \& Hughes 2009).
This study will be conducted in four steps: a) for galaxies belonging to the whole cluster, b) as a function of density, 
c) separately for objects within the different cluster substructures, and d) for different morphological
subcategories (dE; dE,N; dS0; Im; BCD...). 
UV data will also be used to search for star forming regions in 
intergalactic clouds such as those found in the Leo ring (Thilker et al. 2009, Michel-Dansac et al. 2010).
Our survey will provide us with a high galactic latitude deep and wide contiguous field with 
complete wavelength coverage, mostly unaffected by galactic extinction and contamination of foreground stars, 
providing us with a unique dataset for any multifrequency study on background structures.
At the same time, large field UV maps will be compared with infrared maps to identify the diffuse cirrus emission
due to blue scattered light (Cortese et al. 2010).\\

\noindent
The last GUViCS observations were obtained in summer 2010 (see Table \ref{Taball}).
At that time the GALEX team declared the FUV detector officially inoperative. The GUViCS observations have thus been completed only in the NUV band, 
with a $\sim$ 40\% coverage in the FUV (see Figs. \ref{guvics1} and \ref{guvics2}). The Virgo cluster region (12h $<$ RA $<$ 13h; 0$^o$ $<$ dec $<$ 20$^o$)
has been tiled with 133 fields in the FUV band (see Fig. \ref{guvics1}) and 237 in the NUV (Fig. \ref{guvics2}) observed with pointed 
observations (excluding the All sky Imaging Survey, AIS) as listed in Table \ref{Taball}.

 \begin{table*}
\caption{GALEX observations of the Virgo cluster region}
\label{Taball}
{\scriptsize
\[
\begin{tabular}{ccccrr}
\hline
\noalign{\smallskip}
Field		&	R.A.	&	Dec.	&\rm{NUV~integration~time}&\rm{FUV~integration~time}\\ 
		&	(J.2000)&	(J.2000)&(sec)			&	(sec)		\\	
\hline    
GI5\_048014\_GAMA12\_13209	 & 12:00:00.00   & +00:28:01.1 & 1568.2	& 1568.2 \\   
GI5\_048015\_GAMA12\_13208	 & 12:02:01.60   & +01:19:47.8 & 1552.0	& 1552.0 \\   
GI1\_077004\_BDp172428	 	 & 12:03:15.20   & +16:41:10.8 & 2730.1	& 2730.1 \\   
GI3\_103010\_MKW4		 & 12:03:57.70   & +02:13:48.0 & 2212.1	& 2212.1 \\   
GI5\_048017\_GAMA12\_13260	 & 12:04:03.97   & +00:29:07.1 & 1537.2	& 1537.2 \\   
GI4\_042028\_AOHI120446p103742   & 12:04:04.44   & +10:48:37.5 & 1213.2 & 1213.2 \\   
GI1\_079001\_NGC4064  		 & 12:04:11.18   & +18:26:36.3 & 3369.9	& 1691.0 \\   
MISDR1\_13206\_0516		 & 12:06:04.28   & +03:03:00.5 & 1557.1	& -	\\  
GI5\_048019\_GAMA12\_13259	 & 12:06:05.51   & +01:20:35.3 & 2990.2	& 2990.2 \\   
GI6\_001094\_GUVICS094\_0001 	 & 12:06:48.00   & +14:21:00.0 & 1668.6 &   -  \\      
GI6\_001001\_GUVICS001		 & 12:07:59.58   & +12:31:28.7 & 1820.0	& -	\\  
MISDR1\_13205\_0517		 & 12:08:05.54   & +03:54:25.8 & 208.0	& -	  \\
MISDR1\_13258\_0517		 & 12:08:06.83   & +02:11:56.3 & 3221.2	& 1544.1 \\   
GI5\_048021\_GAMA12\_13312	 & 12:08:08.97   & +00:29:56.6 & 1677.0	& 1677.0 \\   
GI6\_001002\_GUVICS002		 & 12:08:24.09   & +13:14:13.4 & 2615.5	& -	 \\ 
GI6\_001003\_GUVICS003\_0001 	 & 12:09:36.00   & +15:18:00.0 & 1614.0 &   -  \\      
GI6\_001004\_GUVICS004\_0001 	 & 12:09:48.00   & +14:09:00.0 & 1694.2 &   -  \\      
GI1\_056011\_NGC4147  		 & 12:10:06.20   & +18:32:31.0 & 1678.0	& 1678.0 \\   
MISDR1\_13257\_0517		 & 12:10:08.17   & +03:03:08.8 & 223.0	& -	 \\ 
MISDR1\_13311\_0287		 & 12:10:10.23   & +01:21:04.2 & 3370.4	& 1693.2 \\   
GI6\_001005\_GUVICS005		 & 12:10:12.54   & +11:40:00.5 & 1790.2	& -	\\  
GI6\_001006\_GUVICS006\_0001 	 & 12:11:19.20   & +06:30:00.0 & 1642.1 &   -  \\      
GI6\_001007\_GUVICS007\_0001 	 & 12:11:31.20   & +16:04:48.0 & 1693.0 &   -  \\      
GI2\_125022\_AGESstrip1\_22  	 & 12:11:36.00   & +08:48:00.0 & 3130.0 &   1594.0 \\   
GI6\_001008\_GUVICS008		 & 12:11:36.35   & +05:41:13.4 & 1866.2	& -	 \\ 
GI4\_019003\_SDSS1212p01	 & 12:12:09.31   & +01:36:27.7 &13171.0 &  13171.0 \\  
MISDR1\_13310\_0517		 & 12:12:11.48   & +02:12:04.6 & 3099.3	& 1687.2 \\   
GI5\_048023\_GAMA12\_13365	 & 12:12:14.82   & +00:30:25.5 & 1690.2	& 1690.2 \\   
GI6\_001011\_GUVICS011\_0001 	 & 12:12:17.21   & +12:31:41.5 & 1700.0 &   -  \\      
NGA\_NGC4168			 & 12:12:18.48   & +13:12:35.5 & 2066.0	& -	\\  
NGA\_NGC4192			 & 12:12:22.80   & +14:54:17.2 & 1226.1	& -	\\  
GI1\_079002\_NGC4178  		 & 12:12:46.57   & +11:10:06.9 & 3499.5	& 2231.3 \\   
GI2\_125021\_AGESstrip1\_21  	 & 12:12:48.00   & +08:00:00.0 & 3287.2 &   1597.0 \\   
GI1\_079003\_NGC4189  		 & 12:13:47.26   & +13:25:29.3 & 4865.4	& 1538.0 \\   
GI5\_038015\_IRASF12112p0305 	 & 12:13:47.30   & +02:48:34.0 & 3701.2 &   3701.2 \\   
GI6\_001012\_GUVICS012\_0001 	 & 12:14:00.00   & +15:39:00.0 & 1690.1 &   -  \\      
GI2\_125020\_AGESstrip1\_20  	 & 12:14:00.00   & +07:06:00.0 & 3034.3 &   1520.2 \\   
GI6\_001013\_GUVICS013		 & 12:14:01.72   & +09:54:26.9 & 702.4	& -	  \\
GI2\_007006\_S\_121410p140127	 & 12:14:10.00   & +14:01:27.0 &24304.0 &  15983.8 \\  
GI5\_048025\_GAMA12\_13364	 & 12:14:15.82   & +01:21:10.4 & 1701.0	& 170- \\   
GI6\_001014\_GUVICS014\_0001 	 & 12:14:17.45   & +11:39:12.1 & 1668.2 &   -  \\      
GI6\_001015\_GUVICS015		 & 12:14:40.80   & +03:54:00.0 & 2681.5	& -	\\  
VIRGO\_SPEC\_1		 	 & 12:15:11.00   & +13:06:00.0 & 2604.2	& 1672.0 \\   
GI6\_001016\_GUVICS016\_0001 	 & 12:15:12.00   & +05:30:00.0 & 1631.3 &   -  \\      
GI2\_125019\_AGESstrip1\_19  	 & 12:15:12.00   & +08:48:00.0 & 2930.2 &   1515.2 \\   
GI6\_001017\_GUVICS017\_0001 	 & 12:15:24.00   & +06:18:00.0 & 1657.0 &   -  \\      
GI2\_125018\_AGESstrip1\_18  	 & 12:15:36.00   & +08:00:00.0 & 3160.2 &   1556.2 \\   
GI1\_080028\_MCGp1\_31\_033 	 & 12:16:00.37   & +04:39:03.5 & 3902.5 &   1639.2 \\   
GI6\_001018\_GUVICS018\_0001 	 & 12:16:14.40   & +16:24:00.0 & 1677.0 &   -  \\      
GI6\_001019\_GUVICS019\_0001 	 & 12:16:17.46   & +10:46:37.8 & 1670.2 &   -  \\      
GI5\_048027\_GAMA12\_13419	 & 12:16:21.46   & +00:30:31.6 & 1693.0	& 1693.0 \\   
GI6\_001020\_GUVICS020\_0001 	 & 12:16:22.60   & +12:30:44.5 & 1679.4 &   -  \\      
GI6\_001021\_GUVICS021\_0001 	 & 12:16:48.00   & +14:01:48.0 & 1639.7 &   -  \\      
GI1\_109020\_NGC4235  		 & 12:17:09.90   & +07:11:29.1 & 2897.2	& 1482.1 \\   
GI6\_001022\_GUVICS022\_0001 	 & 12:17:12.00   & +03:06:00.0 & 1689.2 &   -  \\      
GI6\_001023\_GUVICS023\_0001 	 & 12:17:48.00   & +09:15:00.0 & 1475.1 &   -  \\      
GI6\_001023\_GUVICS023\_0002 	 & 12:17:48.00   & +09:15:00.0 & 288.0  &   -  \\      
GI1\_009021\_UGC07332 		 & 12:17:56.70   & +00:27:36.0 & 5396.1	& 1629.0 \\   
GI1\_077009\_Feige59  		 & 12:17:57.60   & +15:38:42.0 & 1565.7	& 1565.7 \\   
GI6\_001024\_GUVICS024		 & 12:18:14.40   & +04:54:00.0 & 1606.2	& -	\\  
GI6\_001025\_GUVICS025\_0001 	 & 12:18:17.50   & +09:54:02.3 & 1706.0 &   -  \\      
GI6\_001026\_GUVICS026\_0001 	 & 12:18:22.16   & +11:38:02.6 & 1612.2 &   -  \\      
GI1\_077001\_BDp192550		 & 12:18:23.90   & +19:08:51.9 & 1645.0	& 1645.0 \\   
GI6\_001027\_GUVICS027\_0001 	 & 12:18:48.00   & +13:24:00.0 & 1696.0 &   -  \\      
GI6\_001028\_GUVICS028\_0001 	 & 12:19:00.00   & +16:33:00.0 & 1680.0 &   -  \\      
GI6\_001029\_GUVICS029\_0001 	 & 12:19:00.00   & +03:52:48.0 & 1684.2 &   -  \\      
GI2\_125016\_AGESstrip1\_16  	 & 12:19:12.00   & +08:00:00.0 & 1482.1 &   1482.1 \\   
GI4\_012001\_PG1216p069		 & 12:19:20.90   & +06:38:38.4 &33699.8 &  30170.3 \\  
GI3\_079021\_NGC4261  		 & 12:19:23.00   & +05:49:31.0 & 1655.0	& 1655.0 \\   
GI2\_017001\_J121754p144525	 & 12:19:24.48   & +14:36:00.0 &27726.0 &  18131.3 \\  
VIRGOHI21			 & 12:19:24.48   & +14:36:00.0 &18571.3 &  -     \\  
GI1\_077007\_HD107227 		 & 12:19:44.80   & +08:40:56.3 & 3051.4	& 1500.3 \\   
GI4\_042074\_J122100p124340	 & 12:19:53.30   & +12:36:41.5 & 1209.0	& 1209.0 \\   
GI2\_125015\_AGESstrip1\_15  	 & 12:20:00.00   & +07:06:00.0 & 6512.6 &   1677.2 \\   
GI6\_001030\_GUVICS030\_0001 	 & 12:20:21.71   & +10:45:15.9 & 1699.0 &   -  \\      
GI6\_001031\_GUVICS031\_0001 	 & 12:20:36.00   & +14:09:36.0 & 1690.0 &   -  \\      
GI6\_001032\_GUVICS032\_0001 	 & 12:20:38.40   & +15:36:00.0 & 1692.0 &   -  \\      
GI6\_001033\_GUVICS033\_0001	 & 12:20:48.00   & +05:18:00.0 & 1668.2 &   -  \\      
GI1\_077008\_FAUSTN\_24		 & 12:20:48.48   & +09:57:14.0 & 3153.6	& 1491.3 \\   
GI1\_079004\_NGC4293  		 & 12:21:12.39   & +18:22:55.7 & 3126.0	& 3126.0 \\   
GI1\_079005\_Group1		 & 12:21:29.14   & +11:30:25.5 & 3281.2	& 1689.2 \\   
GI1\_079006\_Group2		 & 12:21:37.63   & +14:36:06.8 & 1579.3	& 1579.3 \\   
NGA\_NGC4303			 & 12:21:56.14   & +04:28:42.5 & 1992.2	& 941.0 \\    
GI6\_001034\_GUVICS034\_0001 	 & 12:22:00.00   & +03:27:00.0 & 1640.4 &   -  \\      
GI6\_001035\_GUVICS035\_0001 	 & 12:22:00.00   & +09:39:00.0 & 1663.5 &   -  \\      
GI5\_048033\_GAMA12\_13473	 & 12:22:29.44   & +01:20:06.4 & 1644.0	& 1644.0 \\   
GI6\_001036\_GUVICS036\_0001 	 & 12:22:33.60   & +13:12:36.0 & 1692.2 &   -  \\      
GI6\_001037\_GUVICS037\_0001 	 & 12:22:36.00   & +06:18:00.0 & 1678.0 &   -  \\      
GI6\_001038\_GUVICS038		 & 12:22:39.49   & +16:49:01.4 & 2151.1	& -	  \\
GI2\_125013\_AGESstrip1\_13  	 & 12:22:48.00   & +08:00:00.0 & 6552.5 &   1674.1 \\   
\noalign{\smallskip}
\hline
\end{tabular}
\]
}
\end{table*}
\addtocounter{table}{-1}
\begin{table*}
\caption{continue}
\label{Taball}
{\scriptsize
\[
\begin{tabular}{ccccrr}
\hline
\noalign{\smallskip}
Field		&	R.A.	&	Dec.	&\rm{NUV~integration~time}&\rm{FUV~integration~time}\\ 
		&	(J.2000)&	(J.2000)&(sec)			&	(sec)		\\	
\hline
NGA\_NGC4321			 & 12:22:56.17   & +15:49:38.6 & 2932.2	& 1754.1 \\   
GI5\_057001\_NGC4313  		 & 12:22:57.90   & +11:35:04.0 & 3862.1	& 3862.1 \\   
GI5\_057002\_NGC4307  		 & 12:23:10.08   & +08:47:23.0 & 6337.4	& 6337.4 \\   
GI6\_001039\_GUVICS039\_0001 	 & 12:23:36.00   & +14:45:00.0 & 1661.5 &   -  \\      
GI2\_125012\_AGESstrip1\_12  	 & 12:23:36.00   & +07:06:00.0 & 4511.1 &   2939.6 \\   
NGA\_NGC4344			 & 12:23:38.67   & +17:32:45.2 & 1620.2	& 1620.2 \\   
GI1\_079007\_NGC4351  		 & 12:24:01.57   & +12:12:18.2 & 1937.2	& 1937.2 \\   
GI6\_001040\_GUVICS040\_0001 	 & 12:24:21.60   & +05:24:00.0 & 1639.2 &   -  \\      
GI6\_001041\_GUVICS041\_0001 	 & 12:24:26.40   & +10:42:00.0 & 1696.0 &   -  \\      
GI6\_001042\_GUVICS042\_0001 	 & 12:25:00.00   & +03:51:36.0 & 1654.1 &   -  \\      
NGA\_Virgo\_MOS10		 & 12:25:25.20   & +13:10:29.6 & 3128.4	& 1590.2 \\   
GI6\_001043\_GUVICS043\_0001 	 & 12:25:31.20   & +04:41:24.0 & 1490.5 &   -  \\      
GI6\_001043\_GUVICS043\_0002 	 & 12:25:31.20   & +04:41:24.0 & 537.0  &   -  \\      
IRXB\_NGC4385  			 & 12:25:42.81   & +00:34:21.5 & 1662.0	& -	 \\ 
GI2\_125011\_AGESstrip1\_11  	 & 12:25:48.00   & +08:48:00.0 & 2065.2 &   2065.2 \\   
NGA\_Virgo\_MOS08		 & 12:25:49.20   & +11:34:29.6 & 4312.1	& 1602.1 \\   
GI1\_079008\_Group3		 & 12:25:50.59   & +16:06:26.3 & 2152.2	& 2152.2 \\   
VIRGO\_SPEC\_2			 & 12:26:00.00   & +16:07:00.0 & 2415.4	& 2415.4 \\   
GI1\_079009\_Group4		 & 12:26:16.90   & +09:43:07.2 & 2009.2	& 2008.2 \\   
GI6\_001044\_GUVICS044\_0001 	 & 12:26:24.00   & +06:18:00.0 & 1674.1 &   -  \\      
GI2\_125010\_AGESstrip1\_10  	 & 12:26:24.00   & +08:00:00.0 & 3266.4 &   1637.4 \\   
GI6\_001045\_GUVICS045\_0001 	 & 12:26:36.00   & +14:12:00.0 & 1665.2 &   -  \\      
NGA\_NGC4421			 & 12:27:03.61   & +15:27:58.6 & 2052.6	& 1026.5 \\   
GI6\_001046\_GUVICS046\_0001 	 & 12:27:12.00   & +16:45:36.0 & 1524.3 &   -  \\      
GI2\_125009\_AGESstrip1\_09  	 & 12:27:12.00   & +07:06:00.0 & 3188.1 &   1572.0 \\   
NGA\_Virgo\_MOS02		 & 12:27:13.20   & +12:22:29.6 & 2702.2	& 1604.0 \\   
NGA\_NGC4450			 & 12:27:34.46   & +16:42:02.5 & 381.0	& -	 \\ 
GI6\_001047\_GUVICS047		 & 12:27:48.37   & +04:35:13.4 & 1928.0	& -	  \\
GI1\_079012\_Group5		 & 12:27:49.92   & +15:01:23.4 & 1616.1	& 1616.1 \\   
GI3\_089011\_VHC		 & 12:28:07.20   & +01:24:00.0 & 1507.2	& 1507.2 \\   
GI4\_002002\_HI\_1225p01	 & 12:28:07.20   & +01:24:00.0 &21836.2 &  8970.6 \\   
GI6\_001048\_GUVICS048\_0001 	 & 12:28:25.84   & +05:31:32.1 & 1680.0 &   -  \\      
GI6\_001049\_GUVICS049\_0001 	 & 12:28:29.82   & +10:41:10.3 & 1704.0 &   -  \\      
GI1\_077010\_TYC2888961		 & 12:28:46.89   & +07:08:54.7 & 1598.0	& 1598.0 \\   
GI1\_079010\_NGC4457  		 & 12:28:59.00   & +03:34:14.3 & 5723.2	& 2496.0 \\   
GI6\_001050\_GUVICS050		 & 12:28:59.84   & +17:38:13.4 & 2437.3	& -	  \\
GI1\_109010\_NGC4459  		 & 12:29:00.03   & +13:58:42.8 & 1570.8	& 1570.8 \\   
NGA\_Virgo\_MOS09		 & 12:29:01.20   & +13:10:29.6 & 4536.4	& 1403.2 \\   
GI4\_012003\_3C273		 & 12:29:06.70   & +02:19:43.5 &50050.5 &  29948.6 \\  
GI2\_125008\_AGESstrip1\_08  	 & 12:29:24.28   & +08:35:13.4 & 1635.0 &   1635.0 \\   
NGA\_Virgo\_MOS06		 & 12:29:25.20   & +11:34:29.6 & 4451.1	& 1600.0 \\   
GI6\_001051\_GUVICS051\_0001 	 & 12:29:36.00   & +15:54:00.0 & 1375.7 &   -  \\      
GI4\_033001\_SDSSJ122950p020153	 & 12:29:50.58   & +02:01:53.7 & 1658.2 &   1658.2 \\   
GI1\_109011\_NGC4477  		 & 12:30:02.18   & +13:38:11.3 & 1720.4	& 1720.4 \\   
GI6\_001052\_GUVICS052\_0001 	 & 12:30:24.00   & +07:45:36.0 & 1628.0 &   -  \\      
GI6\_001053\_GUVICS053\_0001 	 & 12:30:27.03   & +06:21:30.8 & 1670.2 &   -  \\      
GI6\_001054\_GUVICS054\_0001 	 & 12:30:29.25   & +09:47:53.8 & 1652.2 &   -  \\      
GI1\_080029\_IRASF12280p0133  	 & 12:30:34.43   & +01:16:24.4 & 7506.9 &   1671.2 \\   
MISDR2\_13586\_0520		 & 12:30:46.03   & +01:17:04.8 & 2595.4	& 847.0 \\    
NGA\_Virgo\_MOS01		 & 12:30:49.20   & +12:22:29.6 & 4685.9	& 1576.1 \\   
GI4\_012004\_RXJ1230d8p0115	 & 12:30:50.00   & +01:15:22.6 &46087.9 &  31246.2 \\  
GI6\_001055\_GUVICS055\_0001 	 & 12:31:16.80   & +04:42:00.0 & 1649.2 &   -  \\      
GI6\_001056\_GUVICS056\_0001 	 & 12:31:24.00   & +14:51:00.0 & 1620.1 &   -  \\      
GI1\_079011\_NGC4498  		 & 12:31:39.49   & +16:48:50.4 & 3041.2	& 3041.2 \\   
GI1\_077011\_TYC8775461		 & 12:32:07.52   & +12:15:50.3 & 1702.0	& 1702.0 \\   
GI6\_001057\_GUVICS057\_0001 	 & 12:32:12.00   & +10:48:00.0 & 1691.0 &   -  \\      
GI4\_042076\_AOHI123223p160130	 & 12:32:23.00   & +16:01:30.0 & 1168.4 &   1168.4 \\   
NGA\_NGC4536			 & 12:32:29.40   & +01:58:20.7 & 1762.0	& 1280.0 \\   
NGA\_Virgo\_MOS11		 & 12:32:37.20   & +13:10:29.6 & 3842.3	& 1581.0 \\   
GI1\_033005\_NGC4517  		 & 12:32:45.59   & +00:06:54.1 & 6402.6	& 1906.3 \\   
GI1\_097006\_NGC4517  		 & 12:32:45.60   & +00:06:54.0 & 5178.3	& 2678.1 \\   
NGA\_Virgo\_MOS05		 & 12:33:01.20   & +11:34:29.6 & 4353.4	& 1567.5 \\   
GI6\_001058\_GUVICS058\_0001 	 & 12:33:04.80   & +17:42:00.0 & 1555.1 &   -  \\      
GI2\_125007\_AGESstrip1\_07  	 & 12:33:12.00   & +07:06:00.0 & 3156.3 &   1665.1 \\   
GI6\_001059\_GUVICS059\_0001 	 & 12:33:12.00   & +08:30:00.0 & 1684.2 &   -  \\      
GI1\_079013\_NGC4522  		 & 12:33:39.60   & +09:10:26.0 & 2496.1	& 2496.1 \\   
GI6\_001060\_GUVICS060		 & 12:33:48.00   & +14:06:00.0 & 1719.6	& -	  \\
GI3\_041007\_NGC4526  		 & 12:34:03.00   & +07:41:57.0 & 1660.2	& 1660.2 \\   
GI6\_001061\_GUVICS061\_0001 	 & 12:34:12.00   & +15:09:00.0 & 1629.0 &   -  \\      
GI1\_079014\_NGC4532  		 & 12:34:19.45   & +06:28:02.2 & 2968.2	& 1343.9 \\   
GI6\_001062\_GUVICS062\_0001 	 & 12:34:24.00   & +10:09:00.0 & 1623.8 &   -  \\      
NGA\_Virgo\_MOS03		 & 12:34:25.20   & +12:22:29.6 & 4738.4	& 1588.2 \\   
GI6\_001064\_GUVICS064\_0001 	 & 12:34:48.00   & +04:36:00.0 & 1639.3 &   -  \\      
GI6\_001063\_GUVICS063\_0001 	 & 12:34:48.00   & +05:28:12.0 & 1670.2 &   -  \\      
GI1\_077012\_TYC8742271		 & 12:35:23.93   & +09:26:56.2 & 2233.1	& 2233.1 \\   
GI6\_001065\_GUVICS065		 & 12:35:36.00   & +16:48:00.0 & 1963.1	& -	  \\
GI2\_125031\_AGESstrip2\_09 	 & 12:36:00.00   & +11:00:00.0 & 1706.0 &   1706.0 \\   
NGA\_Virgo\_MOS12		 & 12:36:13.20   & +13:10:29.6 & 4738.5	& 1594.0 \\   
GI6\_001066\_GUVICS066		 & 12:36:24.00   & +16:06:00.0 & 1625.5	& -	  \\
NGA\_Virgo\_MOS07		 & 12:36:37.20   & +11:34:29.6 & 4495.1	& 1575.1 \\   
GI1\_109013\_NGC4570  		 & 12:36:53.40   & +07:14:47.9 & 1653.3	& 1653.3 \\   
GI2\_034006\_Malin1		 & 12:36:59.30   & +14:19:49.5 & 3154.2	& 3154.2 \\   
NGA\_MALIN1			 & 12:37:00.48   & +14:20:07.7 & 1836.5	& 1836.5 \\   
GI2\_125006\_AGESstrip1\_06 	 & 12:37:12.00   & +08:30:00.0 & 1690.0 &   1690.0 \\   
GI1\_077013\_BDp162394		 & 12:37:14.84   & +15:25:11.9 & 1664.0	& 1664.0 \\   
NGA\_NGC4578			 & 12:37:31.67   & +09:33:36.6 & 33.1	& 33.1 \\	  
GI1\_079015\_NGC4580  		 & 12:37:48.40   & +05:39:14.4 & 3037.1	& 1403.6 \\   
GI2\_125005\_AGESstrip1\_05  	 & 12:38:00.00   & +08:00:00.0 & 1661.0 &   1661.0 \\   
NGA\_Virgo\_MOS04		 & 12:38:01.20   & +12:22:29.6 & 3805.0	& 1609.0 \\   
GI1\_109024\_NGC4596  		 & 12:38:19.44   & +10:10:33.9 & 1661.0	& 1661.0 \\   
\noalign{\smallskip}
\hline
\end{tabular}
\]
}
\end{table*}
\addtocounter{table}{-1}
\begin{table*}
\caption{continue}
\label{Taball}
{\scriptsize
\[
\begin{tabular}{ccccrr}
\hline
\noalign{\smallskip}
Field		&	R.A.	&	Dec.	&\rm{NUV~integration~time}&\rm{FUV~integration~time}\\ 
		&	(J.2000)&	(J.2000)&(sec)			&	(sec)		\\	
\hline
GI6\_001067\_GUVICS067\_0001 	 & 12:38:38.40   & +06:30:00.0 & 1680.0 &   -  \\      
GI6\_001068\_GUVICS068\_0001 	 & 12:39:12.00   & +09:27:00.0 & 1666.2 &   -  \\      
GI6\_001069\_GUVICS069\_0001 	 & 12:39:33.60   & +07:27:00.0 & 1605.0 &   -  \\      
GI2\_125030\_AGESstrip2\_08  	 & 12:39:36.00   & +11:00:00.0 & 1658.0 &   1658.0 \\   
GI6\_001070\_GUVICS070\_0001 	 & 12:39:36.00   & +14:55:12.0 & 1630.2 &   -  \\      
GI6\_001071\_GUVICS071\_0001 	 & 12:39:48.00   & +16:51:00.0 & 1607.7 &   -  \\      
GI6\_001072\_GUVICS072		 & 12:40:00.00   & +13:48:00.0 & 1525.1	& -	  \\
GI2\_125029\_AGESstrip2\_07  	 & 12:40:24.00   & +12:00:00.0 & 1658.1 &   1658.1 \\   
GI6\_001073\_GUVICS073\_0001 	 & 12:40:36.00   & +15:57:00.0 & 1590.4 &   -  \\      
NGA\_NGC4612			 & 12:41:20.37   & +07:40:12.0 & 366.0	& 366.0 \\    
GI6\_001074\_GUVICS074		 & 12:41:24.00   & +13:00:00.0 & 2862.3	& -	  \\
GI6\_001075\_GUVICS075\_0001 	 & 12:42:00.00   & +05:55:48.0 & 1629.0 &   -  \\      
GI3\_041008\_NGC4621  		 & 12:42:02.30   & +11:38:49.0 & 1652.2	& 1652.2 \\   
GI6\_001076\_GUVICS076\_0001 	 & 12:42:24.00   & +07:42:00.0 & 1669.3 &   -  \\      
GI6\_001077\_GUVICS077		 & 12:42:40.80   & +09:22:48.0 & 1589.3	& -	  \\
GI6\_001078\_GUVICS078		 & 12:43:23.98   & +15:18:25.4 & 1491.9	& -	  \\
GI2\_125027\_AGESstrip2\_05  	 & 12:43:36.00   & +12:18:00.0 & 1649.2 &   1649.2 \\   
GI1\_079016\_NGC4651  		 & 12:43:42.63   & +16:23:36.1 & 2588.0	& 2588.0 \\   
GI1\_109003\_NGC4660  		 & 12:44:31.98   & +11:11:25.9 & 3113.2	& 1624.1 \\   
GI6\_001079\_GUVICS079		 & 12:44:45.60   & +17:15:00.0 & 1896.3	& -	  \\
MISDR1\_13700\_0522		 & 12:44:59.52   & +03:36:16.4 & 1014.0	& 1014.0 \\   
GI2\_125002\_AGESstrip1\_02  	 & 12:45:00.31   & +07:29:13.4 & 2263.2 &   2263.2 \\   
GI6\_001080\_GUVICS080\_0001 	 & 12:45:06.00   & +13:37:12.0 & 1578.1 &   -  \\      
GI2\_125003\_AGESstrip1\_03  	 & 12:45:12.00   & +08:30:00.0 & 1696.1 &   1696.1 \\   
GI6\_001081\_GUVICS081\_0001 	 & 12:45:28.80   & +06:37:48.0 & 1690.0 &   -  \\      
GI2\_125026\_AGESstrip2\_04  	 & 12:45:36.00   & +10:18:00.0 & 1651.2 &   1651.2 \\   
GI6\_001082\_GUVICS082\_0001 	 & 12:46:03.60   & +13:01:04.8 & 1669.0 &   -  \\      
GI6\_001083\_GUVICS083		 & 12:46:07.20   & +09:09:00.0 & 1574.0	& -	  \\
GI6\_001084\_GUVICS084		 & 12:47:00.00   & +15:57:00.0 & 2199.4	& -	  \\
MISDR1\_13761\_0522		 & 12:47:10.80   & +02:43:35.7 & 2282.3	& 2282.3 \\   
GI6\_001085\_GUVICS085\_0001 	 & 12:47:12.00   & +15:00:00.0 & 1623.2 &   -  \\      
GI2\_125025\_AGESstrip2\_03  	 & 12:47:24.15   & +12:05:13.4 & 2276.8 &   2276.8 \\   
GI6\_001086\_GUVICS086		 & 12:48:00.00   & +16:48:00.0 & 1741.2	& -	  \\
GI1\_032002\_VCC2062  		 & 12:48:14.00   & +10:59:06.0 & 1661.0	& 1661.0 \\   
GI1\_079017\_NGC4694  		 & 12:48:15.19   & +10:58:57.8 & 1517.2	& 1517.2 \\   
GI1\_079018\_NGC4698  		 & 12:48:22.96   & +08:29:14.1 & 2642.2	& 2642.2 \\   
GI6\_001087\_GUVICS087		 & 12:48:48.00   & +10:30:00.0 & 2000.1	& -	  \\
GI2\_125001\_AGESstrip1\_01  	 & 12:48:48.00   & +07:30:00.0 & 1693.0 &   1693.0 \\   
MISDR1\_13823\_0522		 & 12:49:24.55   & +01:51:05.6 & 1212.2	& 1212.2 \\   
GI6\_001088\_GUVICS088\_0001 	 & 12:49:57.60   & +12:55:33.6 & 1666.1 &   -  \\      
GI5\_002021\_HI1250040p065044 	 & 12:50:04.20   & +06:50:51.0 & 1610.1 &   -  \\      
GI2\_125024\_AGESstrip2\_02  	 & 12:50:24.00   & +12:00:00.0 & 1654.2 &   1654.2 \\   
GI6\_001089\_GUVICS089\_0001 	 & 12:50:48.00   & +09:26:24.0 & 1673.0 &   -  \\      
GI2\_125023\_AGESstrip2\_01  	 & 12:51:12.00   & +10:36:00.0 & 1658.1 &   1658.1 \\   
MISDR1\_13822\_0522		 & 12:51:20.99   & +02:38:23.9 & 1174.1	& 1183.1 \\   
GI6\_001090\_GUVICS090\_0001 	 & 12:51:31.20   & +14:43:48.0 & 1671.0 &   -  \\      
MISDR1\_13886\_0292		 & 12:51:40.70   & +00:58:43.2 & 1570.2	& 1570.2 \\   
GI6\_001091\_GUVICS091\_0001 	 & 12:51:48.00   & +13:36:00.0 & 1655.2 &   -  \\      
MISDR1\_13821\_0523		 & 12:53:17.40   & +03:25:42.0 & 1240.5	& 1240.5 \\   
MISDR1\_13885\_0523		 & 12:53:36.15   & +01:45:37.4 & 1246.4	& 1246.4 \\   
GI6\_001092\_GUVICS092\_0001 	 & 12:53:48.00   & +10:15:36.0 & 1611.3 &   -  \\      
GI6\_001093\_GUVICS093		 & 12:53:48.15   & +12:05:13.4 & 2662.5	& -	  \\
GI1\_009046\_UGC08041 	 	 & 12:55:09.00   & +00:09:33.0 & 3034.1	& 1674.1 \\   
GI1\_052002\_KIG557		 & 12:55:16.60   & +00:14:49.0 &15319.1 &  7048.1 \\   
MISDR1\_13884\_0523		 & 12:55:31.58   & +02:32:28.8 & 2497.3	& 2497.3 \\   
MISDR1\_13883\_0523		 & 12:57:27.31   & +03:19:19.7 & 1600.0	& -	  \\
GI3\_084072\_J125942p105420	 & 12:59:02.40   & +10:54:20.0 & 896.0	& 896.0 \\    
\noalign{\smallskip}
\hline
\end{tabular}
\]
}
\end{table*}

\section{UV images}

\begin{figure*}
   \centering
   \begin{tabular}{c}
   \includegraphics[width=10cm]{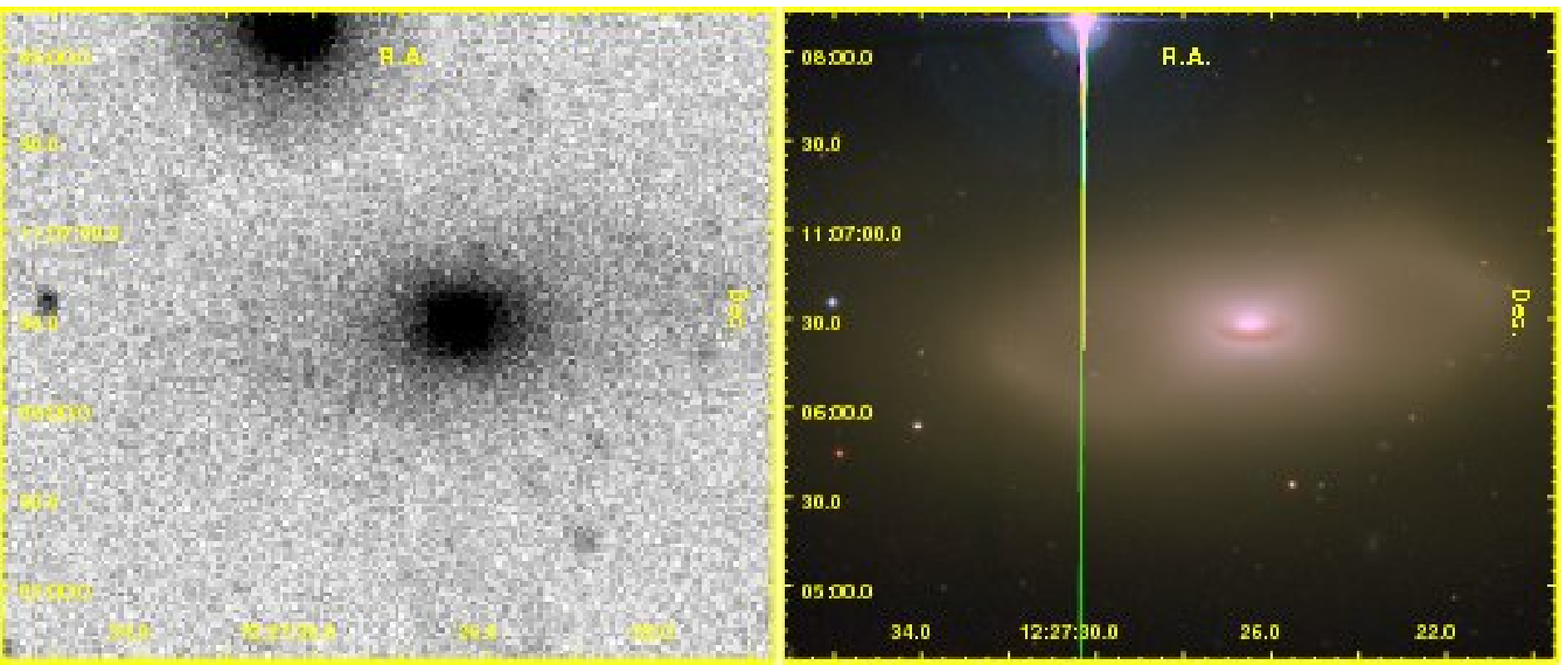}\\
   \includegraphics[width=10cm]{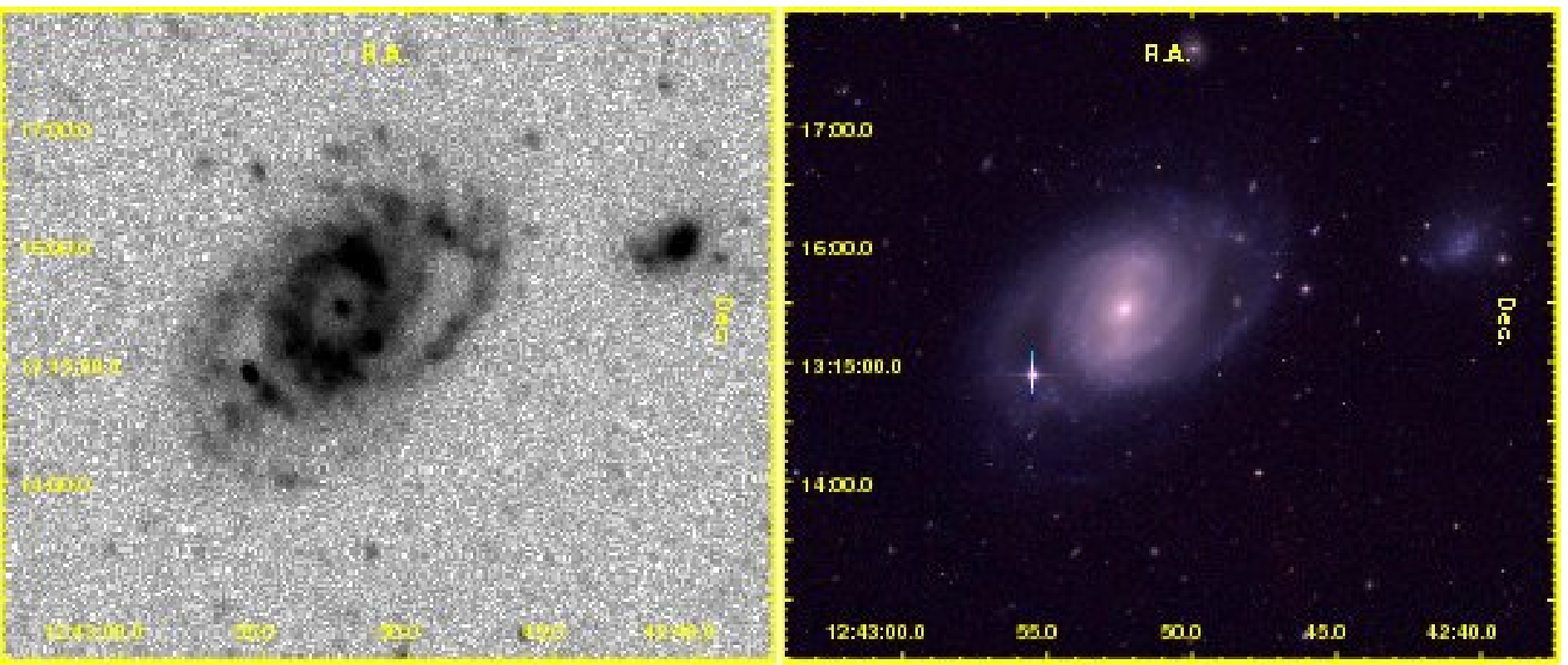}\\
   \includegraphics[width=10cm]{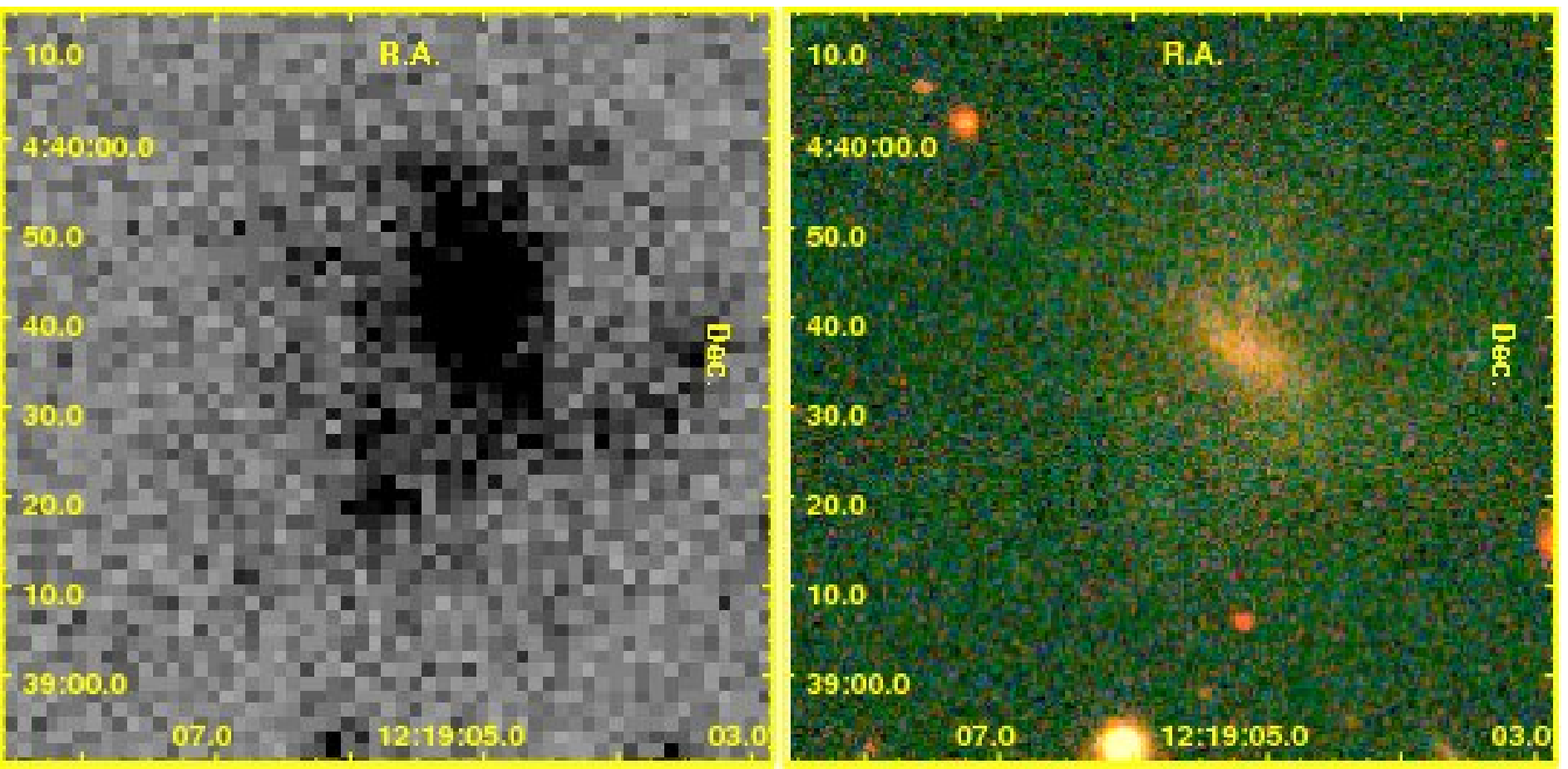}\\
   \includegraphics[width=10cm]{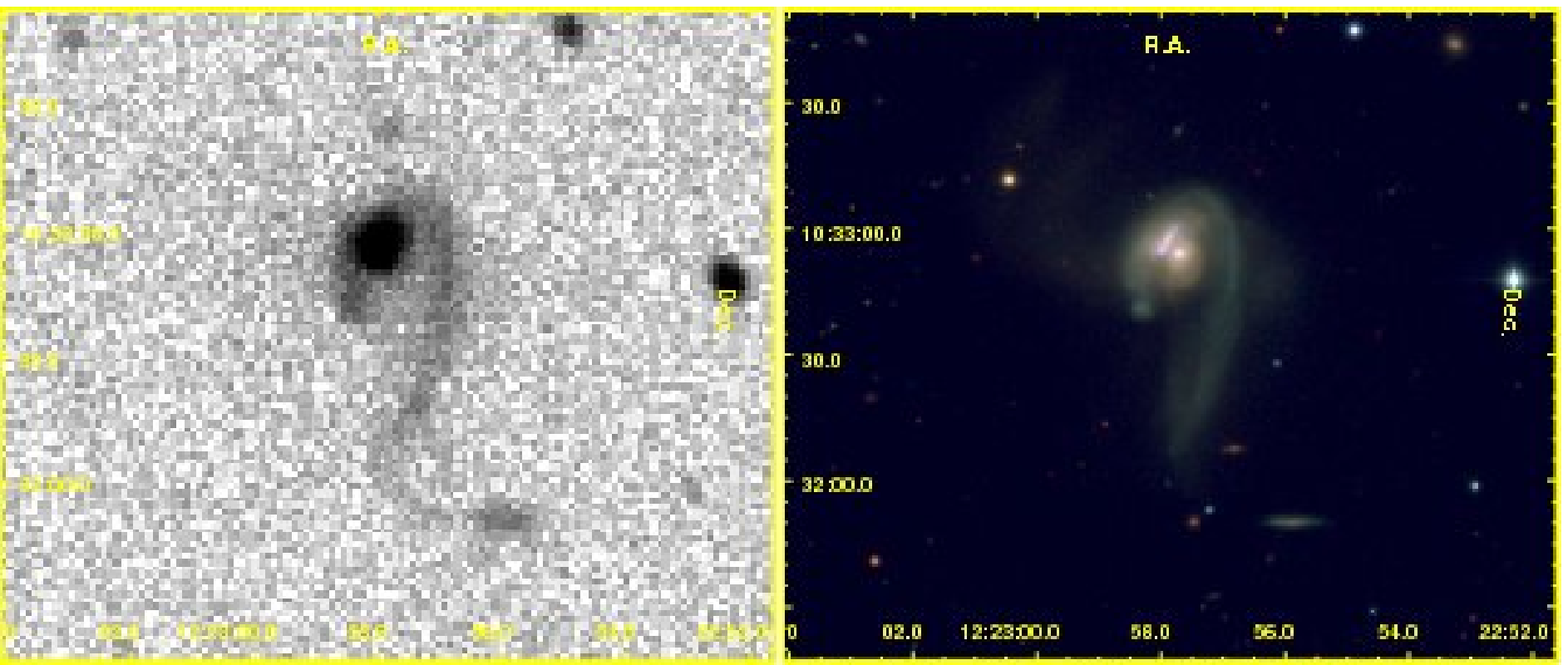}\\
   \includegraphics[width=10cm]{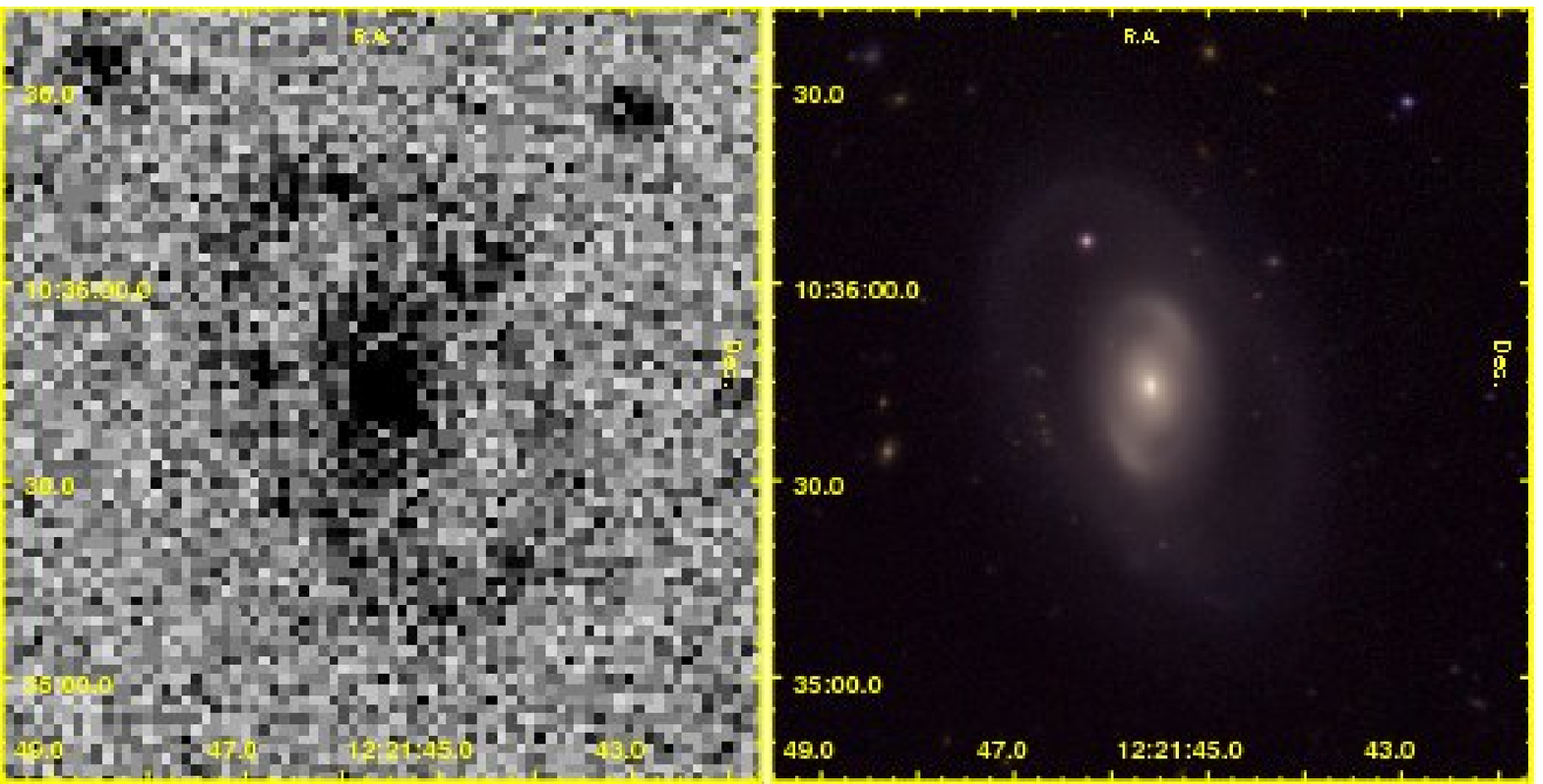}\\
   \end{tabular}
   \caption{From top to bottom, GUViCS NUV (left) and NGVS (SDSS in the case of VCC 320, which is located outside the NGVS footprint; right) $gri$ colour 
   images of a) the lenticular galaxy NGC 4429, b) of the spiral NGC 4639 and of the Im VCC 1931,
   c) of the newly classified Im VCC 320, d) of the background interacting system NGC 4320 and e) of the spiral ring galaxy IC 3199.}
   \label{images}%
   \end{figure*}

The GUViCS survey is providing us with high quality NUV images of different kind of galaxies of sensitivity and angular resolution
well adapted for fulfilling the different scientific purposes of this project. This is well depicted in Fig. \ref{images} representing the NUV GUViCS (left)
and optical NGVS or SDSS (right) images of several Virgo cluster and background objects.

Early-type galaxies, such as the lenticular (S0A+(r)) NGC 4429, can be resolved in the NUV band, thus permitting a detailed study of the radial variation of the 
different stellar populations contributing to the UV emission in evolved systems. The distribution of the young stellar populations produced in recent star formation
events is well traced by the NUV images of both giant and dwarf late-type Virgo cluster members, as shown in the images of the spiral galaxy NGC 4639 (SAB(rs)bc) and of its close companion 
VCC 1931 (Im) or of the late-type VCC 320 (S?). For the latter, the optical image did not allow an accurate morphological classification 
even using the excellent Du Pont photographic plates (Binggeli et al. 1985), however the NUV image undoubtedly identifies it as a star forming dwarf (Im). GUViCS provides us also 
with interesting data for background objects, as in the case of the interacting system NGC 4320, where the NUV image can be used to study the star formation history of the pronounced tidal tails, 
or in IC 3199, a barred spiral with a prominent UV ring barely detected in the optical image.

\begin{figure*}
   \centering
   \includegraphics[width=12cm]{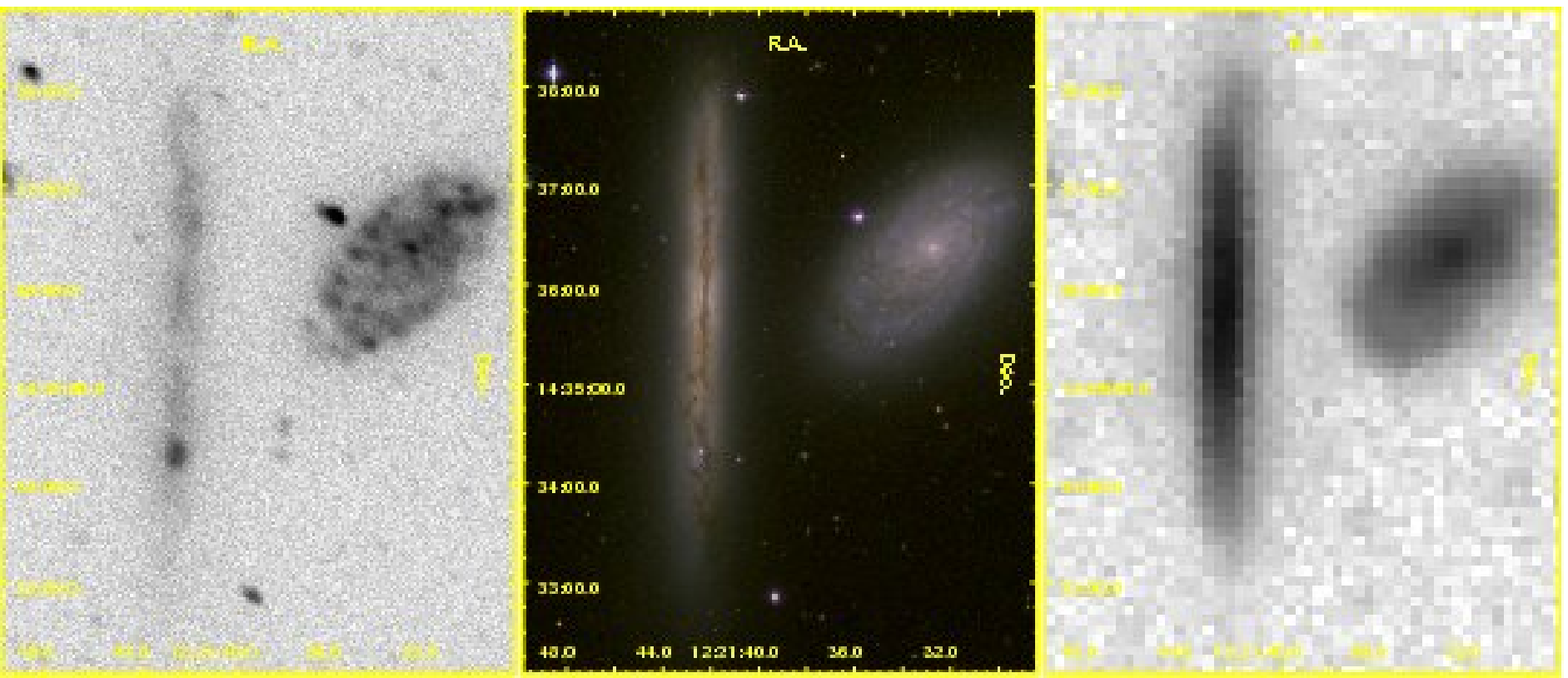}
   \caption{The NUV (GUViCS; left), optical $gri$ (NGVS; middle) and far infrared (HeViCS, 250 $\mu$m; right) images
   of the face-on NGC 4298 and the edge-on NGC 4302 spiral galaxies in the Virgo cluster.}
   \label{taglio}%
   \end{figure*}

The combination of UV (GALEX), optical (NGVS), far infrared (HeViCS) and HI (ALFALFA) data of the whole Virgo cluster region will provide us with a unique dataset 
for a complete and coherent study of all kind of galaxies in different environments. An indicative example of the quality of the science that the unified, multifrequency dataset can provide 
is given in Fig. \ref{taglio} showing the NUV (left), optical (center) and far infrared (right) images of the nearly face-on spiral galaxy 
NGC 4298 (SA(rs)c) and the edge-on spiral NGC 4302 (Sc).
The optical image shows a prominent dust lane in NGC 4302 absorbing almost all the stellar light emitted along the plane of the disc of the galaxy. Because of the high attenuation of 
the ultraviolet light due to the edge-on projection of the galaxy, the UV surface brightness of NGC 4302 is much lower than that of the less extinguished face-on NGC 4298, while the opposite is true 
in the optical bands. The absorbed ultraviolet stellar radiation heats the dust located along the edge-on disc of NGC 4302 and is re-emitted in the far infrared, 
causing the galaxy to have high surface brightness at these wavelengths.
The combination of multifrequency data with radiative transfer models of galaxy discs of different geometries will allow us to quantify the internal attenuation of the UV and optical 
light within these objects and thus quantify the relative weight of the different stellar components to their stellar emission.

\section{Data reduction}

\begin{figure}
   \centering
   \includegraphics[width=9.5cm]{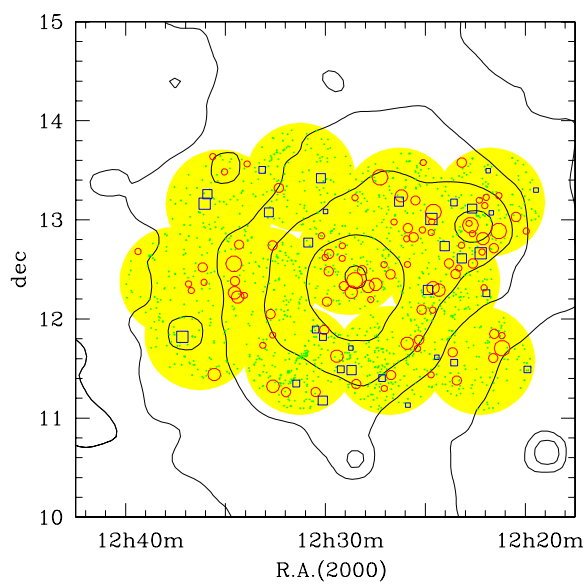}
   \caption{The sky distribution of the GUViCs sources described in this work: large red empty circles and blue squares indicates early- (E, S0, S0a, dE) and late-type
   (Sa-Im-BCD) VCC galaxies respectively detected in at least one UV band. Their size decreases according to their optical photographic magnitude (taken 
   from the VCC Catalogue) in four different bins in magnitude: $<$ 12, 12 $\leq$ mag $<$ 14, 14 $\leq$ mag $<$ 16 and $\geq$ 16. The green dots indicate
   GALEX standard pipeline detections of objects identified as galaxies in the SDSS with NUV magnitudes brighter than 21 mag and include Virgo members and background objects.
   The black solid contours show the X-ray gas distribution obtained with ROSAT (Boheringer et al. 1994).
   }
   \label{GUViCSdetections}%
   \end{figure}

The GALEX pipeline provides images and magnitudes for all of the observed fields. This pipeline gives magnitudes 
extracted using SExtractor (Bertin and Arnouts 1996), as extensively
described in Morrissey et al. (2007). This automatic procedure for flux extraction is optimized for point-like sources within MIS fields, and is thus ideal
for the background objects that dominate the cluster fields. It generally fails for extended sources such as the
bright Virgo galaxies. Different flux extraction procedures must thus be 
adopted for point-like and extended sources.\\
As the first paper of the GUViCS survey, here we briefly present and discuss a flux extraction technique adapted for extended sources. As 
a scientific case we apply it to the central 12 sq. deg. of the cluster, one of the few regions with available FUV and NUV data (see Fig. \ref{GUViCSdetections}). These data are 
then used to determine the FUV and NUV luminosity functions of the core of the Virgo cluster. 
The central 12 sq. deg. of Virgo have been observed at $\sim$ 1500 sec per field in the FUV and $\sim$ 3000-4500 in NUV by the GALEX team
in 2004-2005 (NGA\_Virgo fields in Table \ref{Taball})\footnote{
The central 12 deg$^2$ fields have been monitored since 2005 to search for supernovae: combining the the different images, the total integration
time per pointing is of the order of 15000 sec. These longer exposures are however still not available on MAST.}. 
The GR6 data analyzed in this work have been processed using the GALEX pipeline version Ops7.0.1.

   \begin{figure}
   \centering
   \includegraphics[width=12cm]{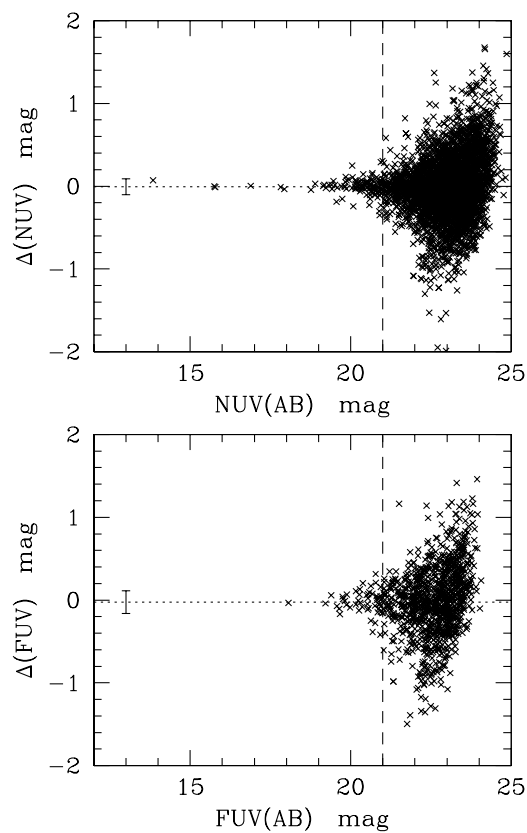}
   \caption{The comparison of NUV (upper panel) and FUV (lower panel) magnitudes obtained form the standard GALEX pipeline for galaxies detected 
   in more than one field. The vertical dashed line indicates the limiting magnitude (21) adopted in this study. The error bar on the left gives the
   1 $\sigma$ dispersion in the data, while the horizontal dotted line gives the mean value, both determined for galaxies with $<$ 21 mag.}
   \label{GALEXcomp}%
   \end{figure}

\subsection{Point-like sources}

The pipeline gives 96207 and 21349 1-sigma detections in the NUV and FUV fields respectively\footnote{Given the poor angular resolution of GALEX, $\sim$ 4-5 arcsec, 
these numbers include stars and artifacts.}. Since the purpose of the present paper 
is to study the UV luminosity function of the Virgo cluster, the flux extraction must be limited to within the completeness magnitude range
of the survey.
The GALEX MIS is complete down to FUV and NUV $\sim$ 21.5 (Xu et al. 2005), we thus adopt a conservative limiting magnitude of 21 mag in both the NUV and FUV bands.
The flux limited sample includes 3467 and 834 sources in the NUV and FUV bands respectively.
The overlap regions in contiguous fields will be used for an accurate calibration of the data and 
for a realistic quantification of the uncertainties (see Fig. \ref{GALEXcomp}).\\
At this magnitude limit, the difference between the fluxes of galaxies observed on more than one frame is $\Delta (FUV)$ = -0.02 $\pm$ 0.14 mag
and  $\Delta (NUV)$ = 0.00 $\pm$ 0.10 mag (see Fig. \ref{GALEXcomp}).\\

\subsection{Extended sources}

\begin{figure*}
   \centering
   \includegraphics[width=12cm]{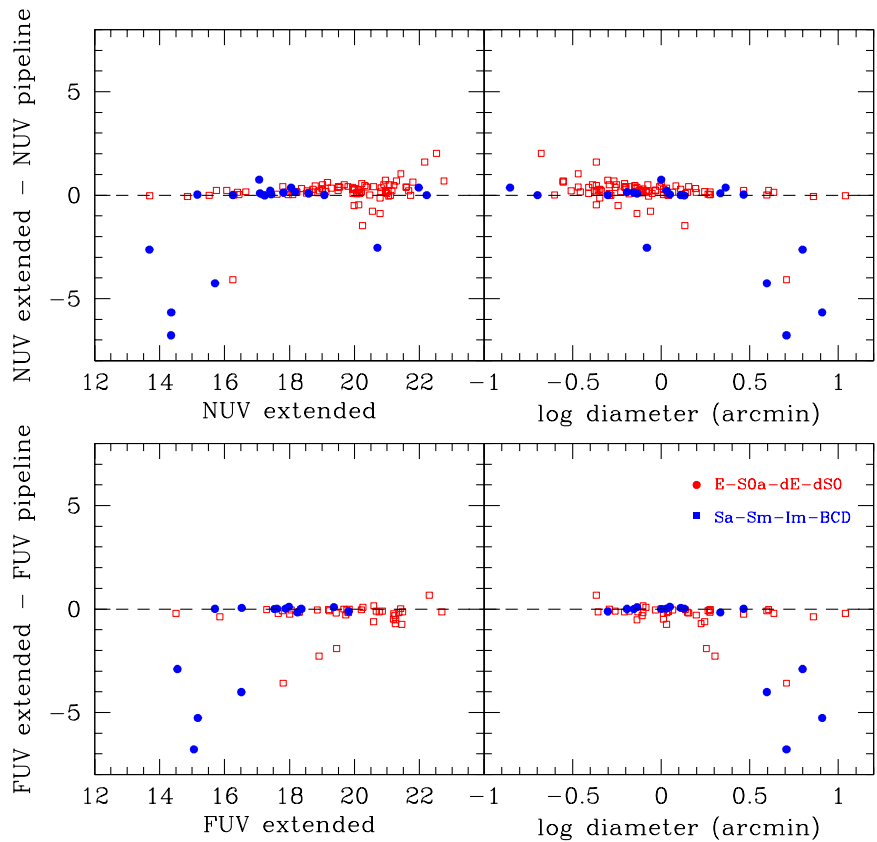}
   \caption{Difference between the NUV (upper panel) and FUV (lower panel) magnitudes obtained using the DS9 FUNTOOLS task
   (extended) and those obtained from the GALEX standard pipeline (pipeline) as a function of the total extended magnitude (left) and the optical diameter (right) of VCC galaxies.
   Open red squares are for early-type galaxies, blue filled circles for late-type objects.}
      \label{compFTpoint}%
   \end{figure*}

Extended sources such as Virgo cluster galaxies, given their proximity (16.5 Mpc) have angular sizes which can reach $\sim$ 10 arcmin. They 
are generally shredded by standard pipelines for flux extraction and must be treated following specially tailored procedures.
This issue is important in giant star forming spirals where compact regions such as nuclei, features 
along the spiral arms and HII regions can be mistaken as individual sources, but the same problem can occur also in low luminosity objects which, except BCDs, still can have
angular sizes of $\sim$ 1 arcmin. To avoid this problem we use the FUNTOOLS task on DS9 to extract the fluxes of all detected VCC galaxies
within the observed frames (325 galaxies). FUNTOOLS allows us to define circular and elliptical apertures and annuli which can be defined, oriented and positioned 
to perfectly cover the image of the galaxy and select a nearby region for the determination of the background sky emission. This technique, although still very crude
since not able to provide us with total extrapolated magnitudes, is well tuned for extended sources. The results can be compared to the the observed fluxes determined with 
SExtractor by the pipeline tailored for point-like sources. In particular this technique is perfectly suited for 
the purpose of the present paper, i.e. the determination of the luminosity function of the Virgo cluster, which is based on histograms with bins that are 
one magnitude wide (the mean difference between extrapolated and aperture magnitudes is generally $\sim$ 0.1 mag in the optical and near infrared bands 
(Gavazzi et al. (2000), a value that might slightly change in the UV bands\footnote{
An accurate description of the flux extraction technique in Virgo cluster galaxies based on the extrapolation of light profiles, with the determination
of total magnitudes and various structural parameters (effective surface brightness and radii, concentration indices) will be presented in a forthcoming 
paper.}). In the present work, FUNTOOLS regions (inclination and position angle) have been defined based on the optical images of the VCC galaxies. Fluxes have been obtained 
integrating the UV images down to the VCC optical diameter which roughly corresponds to an optical surface brightness $\Sigma(B)$ = 25 mag arcsec$^{-2}$. \\
Figure \ref{compFTpoint} compares the integrated magnitudes of VCC galaxies
measured with FUNTOOLS (extended) to those automatically extracted by the GALEX pipeline (pipeline) for early- (red open squares) and late-type (blue filled
circles) galaxies. Figure \ref{compFTpoint} clearly shows that shredding\footnote{The authomatic pipeline resolves different features (nucleus, HII
regions, spiral arms...) in the image of an extended object and consider them as independent galaxies of low luminosity. }
is very important in the most luminous or extended galaxies, where magnitudes from the standard pipeline
can be underestimated by up to $\sim$ 5 mag while, as expected, the effect is much less important in low luminosity and compact systems. This effect is significantly more important 
in late-type systems than in smooth ellipticals because of their complex UV morphology.\\

\subsection{Galaxy - star identification}

The complete NUV and FUV catalogues have been cross-correlated with the SDSS
catalogue (SDSS 7, Abazajian et al. 2009) for the identification of stars and galaxies, which is impossible on UV images because of their poor angular resolution ($\sim$ 4-5 arcsec).
The cross-correlation between GALEX and SDSS used the following criteria. We considered only GALEX objects within 0.5 degrees from the 
center of each GALEX field. We removed duplicates in GALEX overlapping fields by choosing the detections in the field with the largest exposure time. 
The cross correlation between GALEX and SDSS was performed following Budavari et al. (2009), using a positional match radius of 7 arcsec. 
We only considered here GALEX objects with one and only one SDSS counterpart.\\
Out of the 3467 NUV and 834 FUV detected sources with a UV mag brighter than 21 mag, 3333 (NUV) and 779 (FUV) have an optical counterpart in the SDSS.
For the detections in common with SDSS the galaxy-star identification 
is based on SDSS criteria based on the shape of the observed optical PSF (Stoughton et al. 2002), which is quite accurate in the magnitude range
analyzed in this work. Spurious optical identifications are, however, possible for very bright stars for which the optical halo, due to the reflection
of the star light within the telescope, produces an extended image identified as a galaxy. These detections, all inspected by eye, however, are
easily identified as spurious and removed since they are not associated with any bright cataloged galaxy in the VCC. For the 134 objects
in NUV and 55 objects in FUV not having a SDSS counterpart,
the galaxy-star identification is made by means of the UV surface brightness, which is significantly higher in point-like sources than in extended objects
(see Fig. \ref{sigma} and sect. 5). In summary, once stars and artifacts are removed, the resulting sample is composed of
1336 and 660 galaxies detected in NUV and FUV respectively\footnote{The contamination of stars in the NUV is significantly more important than in the FUV
since the far UV band is sensitive only to very young stars which are rare compared to those of intermediate age detectable in the NUV band}, out of which 88 (NUV) and 70 (FUV) are VCC galaxies.

\subsection{Ultra Compact Dwarf Galaxies}

\begin{figure}
   \centering
   \includegraphics[width=14cm]{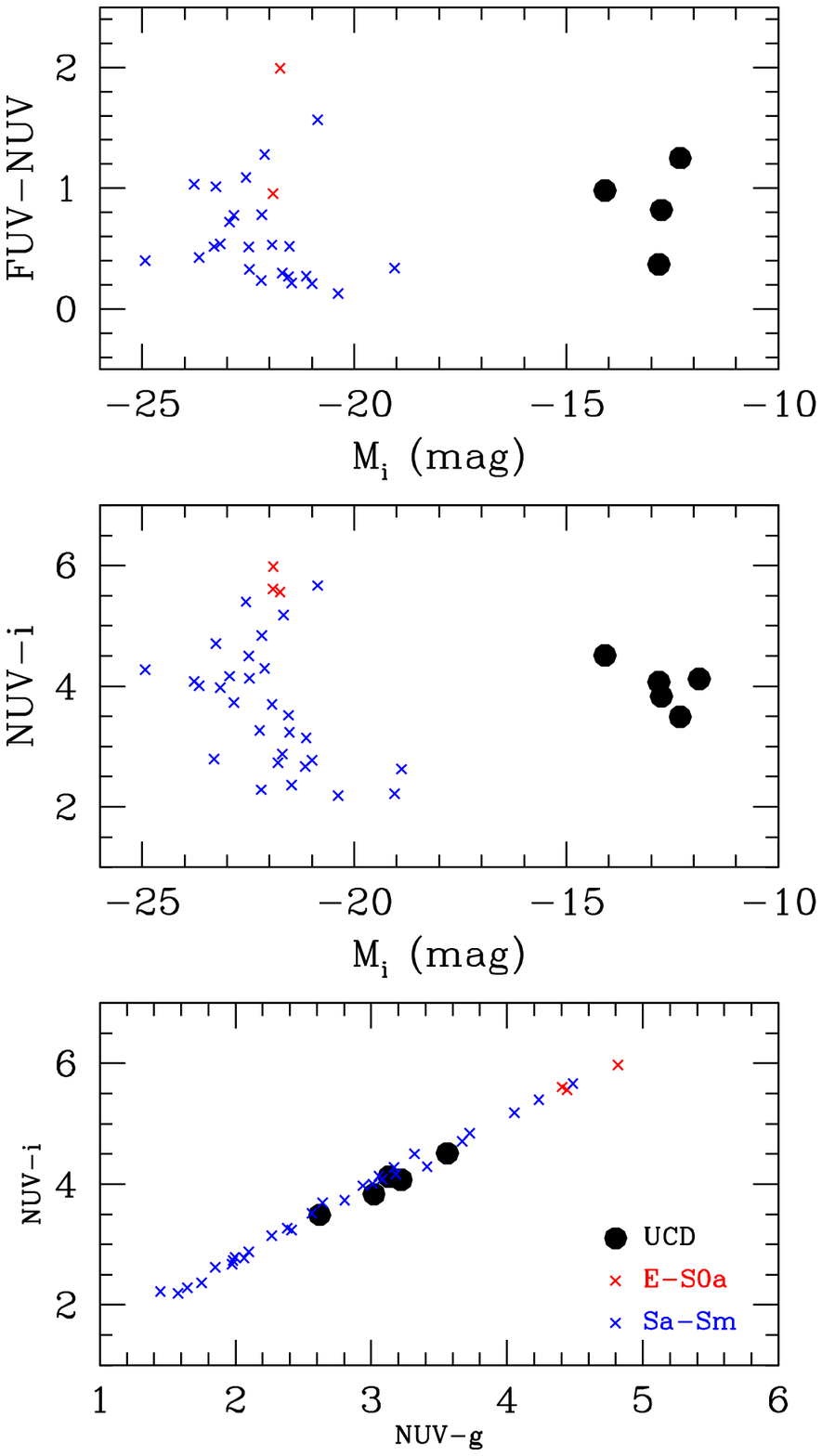}
   \caption{FUV-NUV vs. $M_i$ (upper), NUV-i vs. $M_i$ color-magnitude (middle) and NUV-i vs NUV-g color-color (lower) relations for UCD galaxies (black
   filled dots)
   compared to those observed for normal galaxies of other morphological type extracted from the SINGS sample (from Mu\~noz-Mateos
   et al. 2009; crosses). Red symbols are for early-type galaxies, blue for late-type systems.
   }
   \label{UCD}%
   \end{figure}

Ultra compact dwarf (UCD) galaxies such as those observed by Hasegan et al. (2005) and Jones et al. (2006) are Virgo cluster objects
with a point-like aspect and are thus generally unresolved in ground based optical imaging (Evstigneeva et al. 2008) \footnote{The excellent imaging quality of the NGVS
will allow us to resolve these objects, that will thus be extensively studied in the next future by combining NGVS and GUViCS data}.
The star-galaxy classification based on SDSS optical imaging is thus unable to resolve and identify these objects as galaxies. Virgo UCD galaxies, 
if not identified as cluster members in spectroscopic surveys, are thus not included in the present study. To check whether this selection effect can contaminate 
the determination of our luminosity function we studied the 22 UCD
galaxies observed by Hasegan et al. (2005) and Jones et al. (2006), all located inside the 12 sq. deg. analyzed in this work, and we found that only 6 have been detected by 
GALEX in at least one band. All of them have UV magnitudes below our cutoff limit of 21 mag (see Table \ref{TabUCD}).
Furthermore we notice that these objects do not follow the standard color-magnitude relation of brighter galaxies (see Figure \ref{UCD}) probably because of the presence of a relatively 
old stellar population (Evstigneeva et al. 2007). They have colours similar to those of massive late-type galaxies and are thus not bright objects in the UV bands.
If these 22 objects are representative of the UCD galaxy population of Virgo, we conclude that the missclassification of UCD galaxies as stars does not affect the determination of our UV luminosity functions.

   \begin{figure}
   \centering
   \includegraphics[width=12cm]{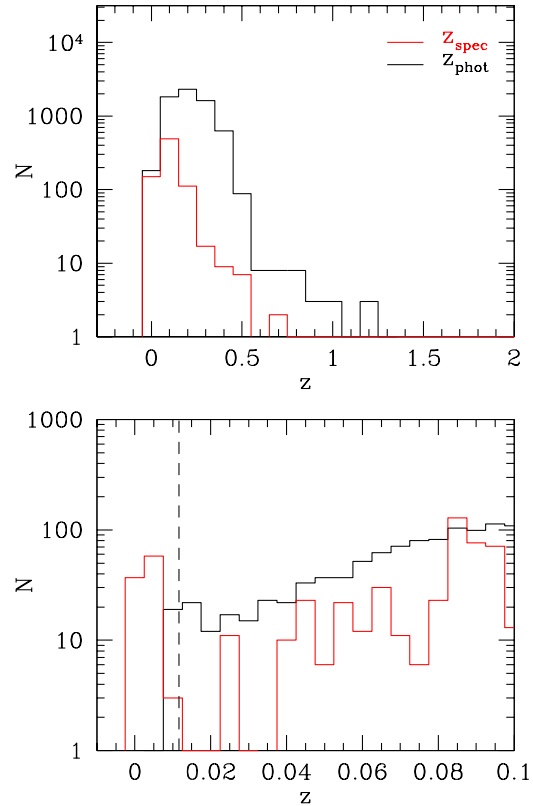}
   \caption{Distribution in photometric (black solid line) and spectroscopic (red solid line) redshift of the UV detected galaxies (upper panel). Photometric resdhifts are
   available for GALEX detected galaxies with a SDSS counterpart with $r$ $<$ 20 mag.
   Zoom on the redshift range 0$<$ $z$ $<$ 0.1 (lower panel). The vertical dashed line shows the adopted limit in the redshift space of the Virgo cluster.}
   \label{zdist}%
   \end{figure}

   \begin{figure*}
   \centering
   \includegraphics[width=16cm]{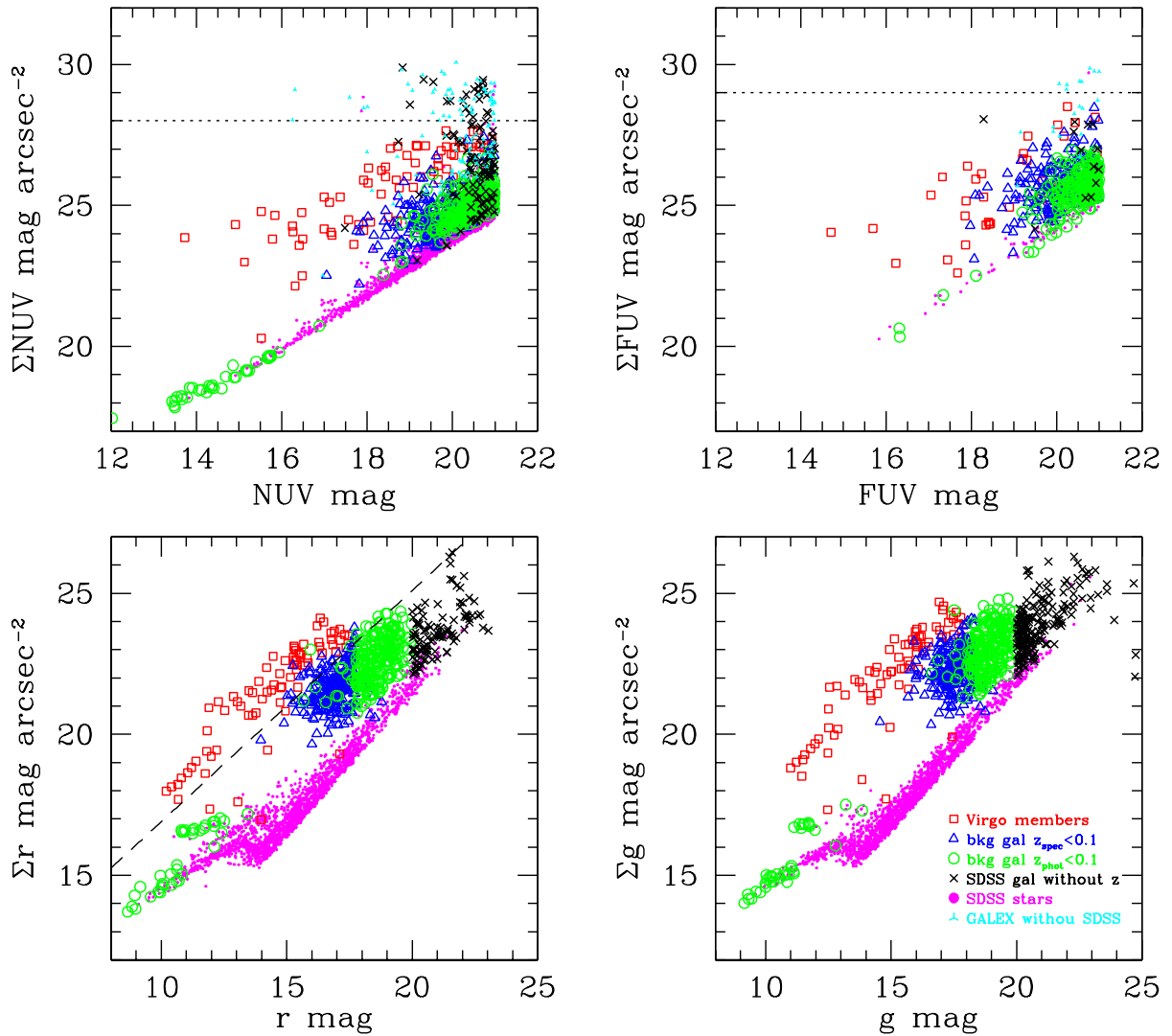}
   \caption{Surface brightness vs. magnitude in four different bands. Red empty squares indicate spectroscopically identified Virgo cluster members, blue empty triangles
   background galaxies with spectroscopic redshift $z_{spec}$ $<$ 0.1, green empty circles objects with photometric redshift $z_{phot}$ $<$ 0.1, black crosses
   galaxies with SDSS data but without spectroscopic or photometric redshift, small magenta dots objects identified as stars in the SDSS and small cyan crosses GALEX detections without 
   SDSS counterparts. The dashed line indicates the surface brightness vs. magnitude limit used for discriminating Virgo members from background objects. The dotted lines indicate
   the limit in surface brightness adopted for selecting galaxies in the UV blind survey.}
   \label{sigma}%
   \end{figure*}

\begin{table*}
\caption{UV and SDSS observed magnitudes for the UCD galaxies detected by GALEX}
\label{TabUCD}
{\scriptsize
\[
\begin{tabular}{cccccccccc}
\hline
\noalign{\smallskip}
Name	&		RA(2000) &	Dec &	FUV	& NUV &	$u$ &	$g$ &	$r$ &	$i$ &	$z$ \\
\hline
S804			&123051.2&	+122613&	21.34&	-	&21.68 & 19.73 &  19.02 &  18.66 &  18.63\\
S1370			&123037.4&	+121918&	-    &	23.25	&22.47 & 20.12 &  19.45 &  19.13 &  18.89\\
J123007.6+123631	&123007.6&	+123631&	22.62&	22.25	&20.49 & 19.03 &  18.45 &  18.18 &  18.11\\
J123048.2+123511	&123048.2&	+123511&	22.89&	22.07	&20.27 & 19.05 &  18.44 &  18.24 &  18.00\\
J123152.9+121559	&123152.9&	+121559&	22.40&	21.42	&19.33 & 17.86 &  17.20 &  16.91 &  16.74\\
J123204.3+122030	&123204.3&	+122030&	23.42&	22.17	&21.27 & 19.55 &  18.99 &  18.68 &  18.55\\

\noalign{\smallskip}
\hline
\end{tabular}
\]
}
Notes: Names are from Hasegan et al. (2005) and Jones et al. (2006); SDSS magnitudes from NED
\end{table*}

\section{The identification of Virgo cluster members}

   \begin{figure*}
   \centering
   \includegraphics[width=16cm]{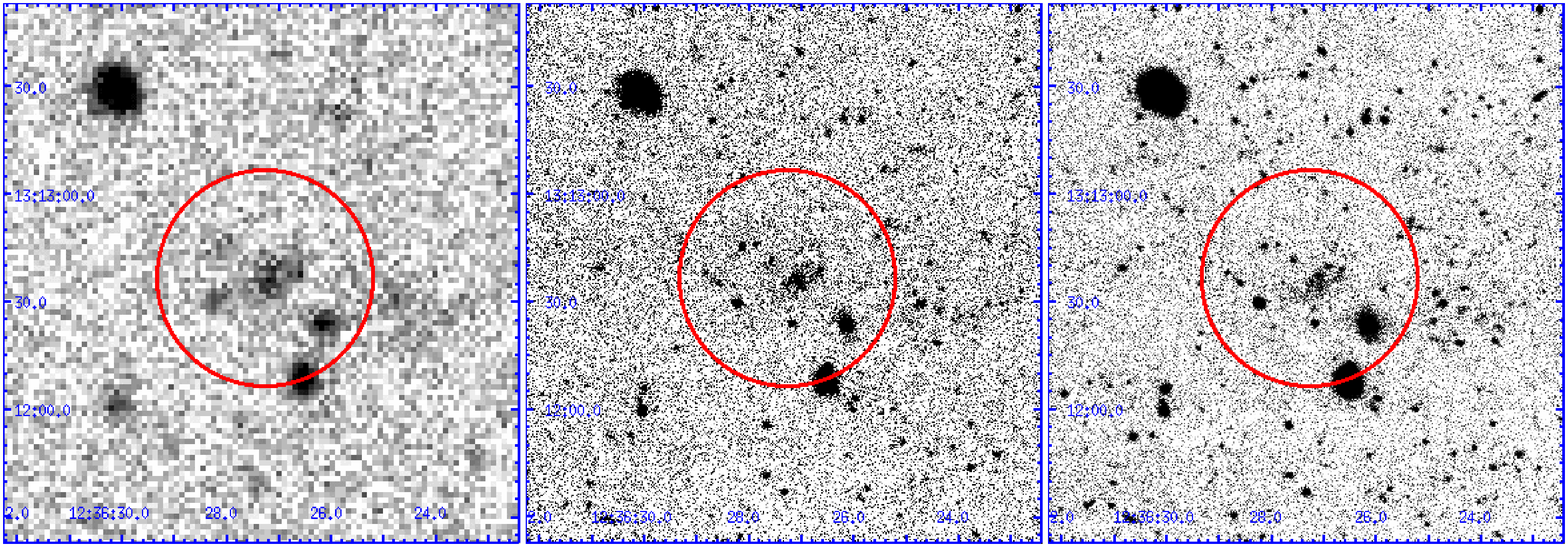}
   \caption{GALEX NUV (left) and NGVS $u$ (middle) and $g$ (right) images of the galaxy GALEX 58202576066067982 undetected by the SDSS..}
   \label{image}%
   \end{figure*}

Virgo cluster members are identified among the NUV (1336) and FUV (660) detected galaxies adopting the following criteria:
we first cross-correlate our catalogue with NED and SDSS to identify galaxies with available spectroscopic redshift (406 with NUV data
and 329 with FUV data; see Fig. \ref{zdist}). Among these we consider as Virgo members those with $z_{spec}$ $<$ 0.01167 ($vel$ $<$ 3500 km s$^{-1}$) to include galaxies belonging to
all of the different substructures of the cluster (Virgo B, M and W clouds, Gavazzi et al. (1999)) or those VCC galaxies defined as Virgo cluster members by Binggeli et al. (1988; 1993).
For the remaining galaxies we first rejected all objects with SDSS photometric redshift $z_{phot}$ $\geq$ 0.1 (this can be done 
only for galaxies with an $r$ band magnitude brighter than 20, Oyaizu et al. 2008), and for the remaining objects 
(galaxies with $z_{phot}$ $<$ 0.1 and $r$ $<$20; $r$ $\geq$ 20; without SDSS
counterpart) we adopt the same surface brightness technique originally and successfully used by Binggeli and collaborators (Binggeli et al. 1985; 1993).

This technique is on purpose slightly modified to deal with galaxies with a SDSS counterpart (black crosses in Fig. \ref{sigma}) or with targets with only GALEX detections (cyan).
SDSS surface brightnesses in the $r$ and $g$ bands are determined using the Petrosian magnitude and half light radii, while in the UV bands using 
SExtractor kron-like magnitudes and half-light radii\footnote{For extended, VCC sources, cluster membership is assigned using redshift, when available, or membership
as determined in Gavazzi et al. (1999).}. 
Figure \ref{sigma} clearly shows that
the star-galaxy identification given in the SDSS works very well: stars have a significantly higher surface brightness than galaxies. The change in slope in the $\Sigma r$ vs $r$
or $\Sigma g$ vs $g$ relation observed in bright objects is due to the saturation of the SDSS detector combined with the ghost formed around bright stars
(Stoughton et al. 2002). Indeed a few bright stars 
($r$ $<$ 14 and NUV $<$ 16) are missclassified as galaxies just because of the presence of an extended ghost in their images.
This surface brightness criterion for discriminating stars from galaxies can be successfully used also in the UV bands. \\
Given the strong relationship between surface brightness and absolute magnitude first noticed by Binggeli et al. (1985), Figure \ref{sigma} can be
used to discriminate Virgo members, characterized by low surface brightness and luminosity, from background sources of similar apparent magnitude
which, because of their distance, are intrinsically luminous, high surface brightness galaxies. All spectroscopically identified Virgo cluster members have $r$ band surface brightness  $\Sigma r$
fainter than $\sim$ 0.82$r$ + 8.7 (dashed line in Fig. \ref{sigma}), with the exception of a few bright sources for which the SDSS magnitude is poorly constrained.
For galaxies without spectroscopic information (green empty circles and black crosses) we thus identify as Virgo members those with $\Sigma r$ $>$ 0.82$r$ + 8.7.
This criterion, however, cannot be applied to GALEX detections without SDSS counterparts (cyan in Fig. \ref{sigma}).
Objects with only GALEX data have UV surface brightnesses generally smaller than those of VCC galaxies (with magnitudes measured using FUNTOOLS).
This is evidence that they are generally spurious detections since the analysis of all the VCC galaxies within the field revealed that only
galaxies with a NUV and FUV surface brightness brighter than 28 and 29 mag arcsec$^{-2}$ can be detected. 
Indeed the visual inspection of the brightest objects with only GALEX data revealed that they are ghosts associated to bright stars. We thus applied a UV surface brightness limit 
to exclude all detections with $\Sigma NUV$ $>$ 28 mag arcsec$^{-2}$ and $\Sigma FUV$ $>$ 29 mag arcsec$^{-2}$. We then visually inspected the GALEX detections 
without SDSS counterparts satisfying these conditions (55 in NUV and 12 in FUV) and found out that they were a) ghosts of bright sources, b) HII regions in the 
tail of the perturbed galaxy IC 3418 (Hester et al. 2010; Fumagalli et al. 2010) or c) background galaxies at small projected angular distance unresolved by the GALEX pipeline.
Only one object detected by GALEX without a SDSS counterpart
can be added to the NUV detections. This object, GALEX 58202576066067982 (R.A.(J2000) = 12h36m27.28s; Dec = +13$^o$12'36.7"), whose image is shown in Fig. \ref{image}, 
is a very blue low surface brightness galaxy undetected in the SDSS and barely
detected in the $u$ and $g$ NGVS bands with UV AB magnitudes FUV = 21.64 and NUV = 20.95 mag.
Following these selection criteria we end up with 135 and 65 Virgo galaxies detected in NUV and FUV respectively down to NUV and FUV 21 mag, out of which 92 are early-type (E-S0-S0a)
and 43 late-types (Sa-Sd-Im-BCD, where the morphological type has been taken from the VCC) in NUV and 34 and 31 in FUV. 
It is interesting to note that, even after the advent of recent photometric and spectroscopic surveys such as the SDSS or GALEX, the detected Virgo cluster members
are for the majority (124/135 in NUV and 59/65 in FUV) already catalogued in the Virgo cluster catalogue of Binggeli et al. (1985) based on photographic plates taken with the 2.5 meters Du Pont telescope
at Las Campanas.

\section{The UV luminosity function of the central 12 deg.$^2$}

\begin{figure}
   \centering
   \includegraphics[width=14cm]{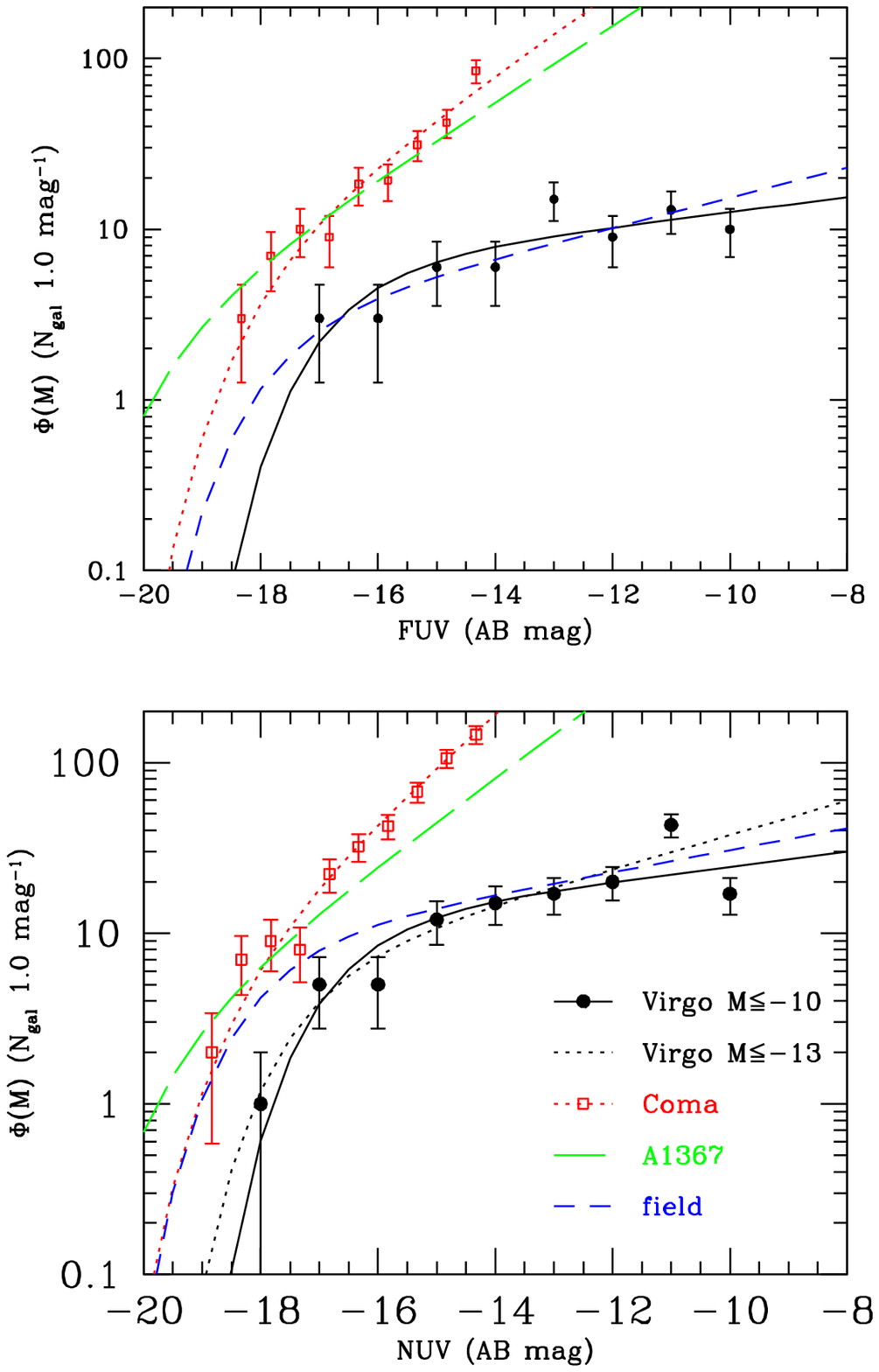}
   \caption{The FUV (upper) and NUV (lower) luminosity functions of the Virgo cluster down to the magnitude limit of $M \leq -10$ (black filled dots; solid line)
   or limited to $M \leq -13$ (black dotted line)
   compared to those obtained for the Coma cluster (red squares; dotted line; from Cortese et al. 2008b), A1367 (green long-dashed line; from Cortese et al. 2005) 
   and the field (blue short-dashed line; from Wyder et al. 2005). The field luminosity function has been normalized to include approximately
   the same number of Virgo objects.
   }
   \label{LF}%
   \end{figure}

\begin{figure*}
   \centering
   \includegraphics[width=16cm]{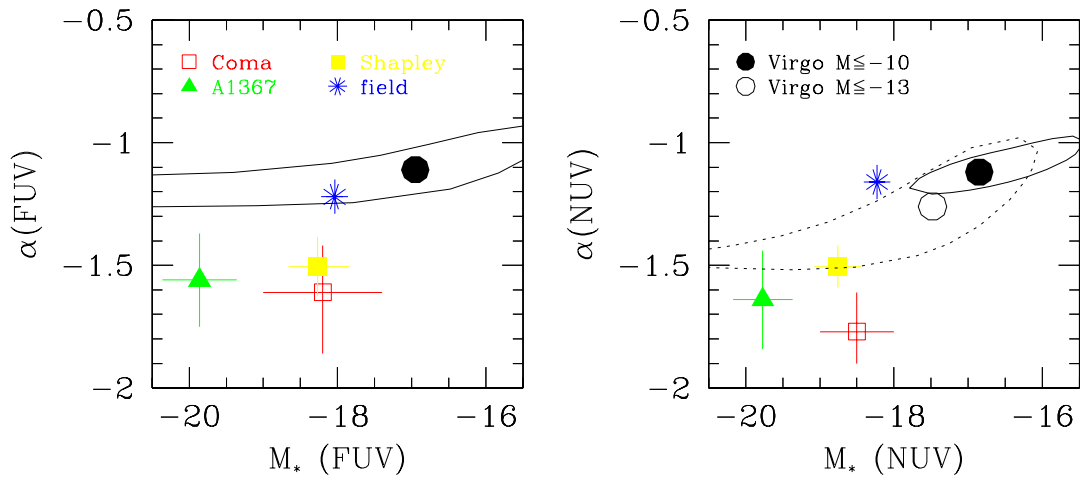}
   \caption{Comparison of the parameters obtained by fitting Schechter functions to the FUV (left) and NUV (right) luminosity functions of the Virgo cluster 
   down to magnitude limits of $M$ $\leq$ -10 (black filled circles and solid contours) and of $M$ $\leq$ -13 (black empty circles and dotted
   contour or error bars) with their uncertainty, of the Coma cluster (red empty square; from Cortese et al. 2008b), of A1367 (green filled triangle; from Cortese et al. 2005),
   of the Shapley supercluster (yellow filled square; from Haines et al. 2010) 
   and of the field (blue asterisk; from Wyder et al. 2005).
   }
   \label{LFpara}%
   \end{figure*}

The FUV and NUV luminosity functions of the central 12 deg.$^2$ can be confidently measured using the available data.
UV magnitudes are corrected for Galactic extinction, here done using the Schlegel et al. (1998) extinction map and the Galactic extinction law of Pei (1992).
The determination of the internal attenuation requires far-infrared data for all the detected sources and is deferred to a future paper.
This correction, however, should be important only for the star forming galaxy population but negligible in the dominant early-type galaxy population,
including dE and transition objects (de Looze et al. 2010).\\
Figure \ref{LF} shows the NUV and FUV luminosity function of the Virgo cluster compared to those obtained for Coma (Cortese et al. 2008b)
and A1367 (Cortese et al. 2005) and for field galaxies (Wyder et al. 2005). The data are fitted with a Schechter function, with parameters given in Table \ref{Tabfit}
and compared in Fig. \ref{LFpara}. For comparison with other clusters we remind that the core and virial radii of Virgo are of $R_C$ = 130 kpc and $R_V$ = 1.68 Mpc (Boselli \& Gavazzi 2006),
corresponding to 0.45$^o$ and 5.8$^o$ respectively at a distance of 16.5 Mpc.

\noindent
As a first remark we notice that, despite the relatively good statistics, the sampled volume is relatively small ($\sim$ 3.4 Mpc$^3$
compared to A1367, $\sim$ 9 Mpc$^3$, and Coma, $\sim$ 50 Mpc$^3$) and thus includes few bright galaxies (the whole GUViCS survey covers the whole
Virgo cluster, whose size is at least 20 Mpc$^3$). For this reason, the value of
$M^*$, which appears to be $\sim$ 1 mag fainter than in the other two clusters or in the field is not well constrained, with a formal uncertainty of $\sim$ 1 mag.
Indeed the values of $M^*$ for Virgo given in Table \ref{Tabfit} are significantly fainter than those obtained using the old FOCA and FAUST surveys that, 
despite being relatively shallow, covered a larger area and therefore included several bright galaxies\footnote{Given the shallow nature of these old surveys, the 
uncertainty on their $M^*$ was mostly due to their limited extension to low luminosities.}. These old surveys have shown that Coma and Virgo 
had similar values of $M^*$, while in A1367 the presence of very active galaxies such as CGCG 97-087, 97-079 and 97-073 shifted $M^*$ to slightly brighter
values (Cortese et al. 2003). To quantify how the lack of bright objects might change our results, we fixed $M^*$ equal to the values obtained for the Coma cluster, -18.50 and -18.20 in the NUV and FUV bands 
respectively, and determined the associated $\alpha$ parameters (see Table \ref{Tabfit}). Although the absolute values of $\alpha$ increase at both wavelengths, 
the corresponding luminosity functions are still significantly flatter than those determined in Coma and A1367.\\
Given the correlation between $\alpha$ and $M^*$, large uncertainties on $M^*$ translate to a significant uncertainty on the $\alpha$
parameter ($\sim$ 0.2). Despite this, the data apperas to indicate
that the UV luminosity functions of the core of the Virgo cluster are significantly flatter ($\alpha$ $\sim$ -1.1) both in the NUV and FUV bands than those of Coma and A1367 
($\alpha$ $\sim$ -1.6) or the Shapley supercluster ($\alpha$ = -1.5; Haines et al. 2010), and more consistent with the luminosity function measured for field galaxies ($\alpha$ $\sim$ -1.2, Wyder et al. 2005).
Several caveats prevent us from making firm conclusions at this stage. First, the Virgo luminosity function presented here extends to fainter magnitudes ($M$ $\sim$ -10) than the Coma, A1367, Shapley 
and even field luminosity functions (which have limiting magnitudes 4.5 mag brighter in the clusters, and 3 mag brighter in the field). If we were to determine the Schechter parameters of the Virgo 
NUV\footnote{The poor statistics prevents to do this exercise in the FUV band.}  
luminosity function limiting the data to $M$ $\leq$ -13 (the required limit of $M$ $\leq$ -14.5 is prohibitive for the poor statistics even in the NUV
band), we would obtain values closer (but still significantly different) to those obtained for the other clusters, although with an even larger error on $M^*$ (see Fig. \ref{LFpara} and Table \ref{Tabfit}).
Second, the Virgo luminosity functions have been measured within a region ($\sim$ 1 Mpc$^2$, $\sim$ 0.35$R_V$) that is less extended than that used to determine the luminosity function in
the Coma cluster ($\sim$ 25 Mpc$^2$ excluding the core of the cluster) but comparable to that of A1367 ($\sim$ 2 Mpc$^2$, $\sim$ 0.37$R_V$). The differences in the measured slopes might therefore be due 
to mass and morphological type
segregation effects. In Coma this effect is compounded by the fact that the luminosity functions have been determined excluding the central $\sim$ 1 sq. deg. region, which could not be observed
because of the presence of bright stars saturating the FUV detector. There are indeed some hints that in Coma the slope of the UV luminosity functions
flattens going from the periphery to the densest observed regions (Cortese et al. 2008). 
Another source of uncertainty in Coma and A1367 is that in the faint regime ($M$ $\geq$ -16) 
cluster membership is determined for the majority of galaxies using poorly constrained statistical criteria
whereas in Virgo more accurate membership can be assigned down to $M$ = -10 (Rines \& Geller 2008).

\begin{table}
\caption{The best fitting parameters for the NUV and FUV luminosity functions}
\label{Tabfit}
{\scriptsize
\[
\begin{tabular}{cccc}
\hline
\noalign{\smallskip}
Band		&	Sample.	&	$M^*$	&$\alpha$\\
\hline
NUV		&	All	& -16.86$^{+0.73}_{-0.62}$	&-1.12$^{+0.07}_{-0.06}$	\\
NUV(M$<$-13)	&	All	& -17.48$^{+0.85}_{-1.21}$	&-1.26$^{+0.16}_{-0.17}$	\\
NUV		&	All	& -18.50 (fixed)  		&-1.23$^{+0.03}_{-0.03}$	\\
NUV		& Early-types	& -16.36$^{+1.03}_{-1.20}$	&-1.17$^{+0.09}_{-0.08}$	\\
NUV		& Late-types	& -17.60$^{+1.03}_{-1.63}$	&-1.00$^{+0.12}_{-0.12}$	\\
\hline
FUV		&	All	& -16.95$^{+1.30}_{-3.56}$	&-1.11$^{+0.10}_{-0.10}$	\\
FUV		&	All	& -18.20 (fixed)		&-1.17$^{+0.06}_{-0.06}$	\\
\hline
\end{tabular}
\]
}
\end{table}

\begin{table}
\caption{The Virgo cluster luminosity functions in the B band}
\label{TabB}
{\scriptsize
\[
\begin{tabular}{ccccc}
\hline
\noalign{\smallskip}
Band	&	Sample.	&	$M^*$	&$\alpha$\\
\hline
B	&	All	& -20.6	&-1.25	\\
B	& Early-types	& -21.4	&-1.40	\\
B	& Late-types	& -19.7	&-0.80	\\
\noalign{\smallskip}
\hline
\end{tabular}
\]
}
Notes: $M^*$ from Sandage et al. (1985) have been transformed into AB system magnitudes and corrected
for a distance modulus of $m-M$ = 31 mag as in the present work.
\end{table}

\begin{figure}
   \centering
   \includegraphics[width=14cm]{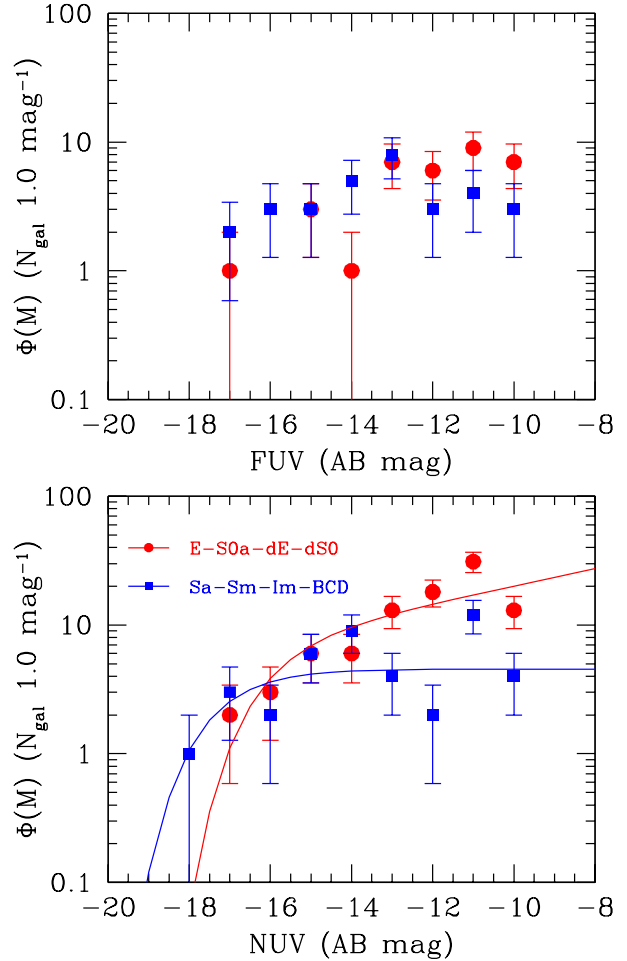}
   \caption{The FUV (upper) and NUV (lower) luminosity functions of early-type (E-S0a-dE-dS0; red symbols) and late-type (Sa-Sm-Im-BCD; blue symbols) Virgo cluster galaxies.
   The red and blue solid lines give the best fit Schechter function for early- and late-type galaxies in the NUV band.
   }
   \label{LFtype}%
   \end{figure}

\noindent
Despite these large uncertainties, we tentatively compare these results to those obtained at other frequencies.
The faint end slope of the NUV and FUV luminosity functions of Virgo galaxies is slightly flatter than that obtained by Sandage et al. (1985)
in the B bands down to the completeness limiting magnitude of their survey ($B_T$ = 18; $\alpha$ = -1.25; see Table \ref{TabB}). 
We remind, however, that the B band luminosity function of Sandage et al. (1985) is determined on the whole Virgo cluster ($\sim$ 10 sq. deg.).
A steepening of the faint end slope of the UV luminosity function in the outskirts of Virgo is expected since already observed 
in the optical bands in other clusters such as Coma (Beijersbergen et al. 2002; Adami et al. 2007, 2008). \\
The accurate morphological classification available for the VCC galaxies (from Binggeli et al. 1985, those not catalogued in the VCC being classified as late-type dwarfs
according to their optical morphology and blue colors) allows us to estimate the UV luminosity function separately for late- (Sa-Sm-Im-BCD) and 
early-type (E-S0a-dE-dS0) systems (see Fig. \ref{LFtype}).

\noindent
In the NUV band the fitted Schechter functions (see Table \ref{Tabfit}) are qualitatively similar to those obtained in the optical band (Table \ref{TabB}), with a flatter slope for late-type systems
($\alpha$ $\sim$ -1.0) than in early-type galaxies ($\alpha$ $\sim$ -1.2) due to the increasing contribution of the dominating dE-dS0 galaxy population at the faint end. The difference
between the slopes of the early- and late-type galaxy populations is however more marked in the optical than in the NUV. This is quite expected since 
dwarf elliptical and spheroidal galaxies have redder colours than dwarf star 
forming systems relatively rare in high density environments.
The difference in $M^*$ between star forming and quiescent systems in the B and NUV bands is consistent with their observed difference in color (Boselli et al. 2005a;
2008a). 
The poor statistics in the FUV data prevents accurate fitting of the data once the sample is divided into star forming and quiescent systems. 
The data are, however, consistent with what found in Coma and A1367, i.e. star forming objects dominate at high luminosities
while quiescent galaxies dominate at low luminosities. We note that the difference between early- and late-type systems will likely be more
pronounced once the luminosity function is determined for the whole Virgo cluster since, given the well known morphological segregation effect,
the core region analyzed in this work is dominated by early-type systems. Furthermore, since gas stripping is expected to be more efficient in the core region, 
the star formation activity, and therefore the FUV fluxes, of the 
late-type galaxies sampled here are likely significantly reduced with respect to those of galaxies
at the periphery of the cluster (Boselli \& Gavazzi 2006). It is interesting to note that, despite the fact that the UV emission is generally an indication
of the presence of very young stellar populations, the UV luminosity function of cluster galaxies is dominated by evolved early-type galaxies.\\
These results on the shape of the UV luminosity functions are qualitatively consistent with the picture recently proposed by Boselli et al. (2008a, 2008b), 
i.e. that low-luminosity star forming galaxies recently accreted by
the Virgo cluster interacted with the hostile environment losing most of their gas after a relatively short ram-pressure stripping event. The lack of gas, the principal feeder of 
star formation, induces a rapid reduction of the star formation activity, and therefore of the UV luminosity (Boselli et al. 2009), transforming star forming rotating systems into
quiescent dwarf ellipticals. Indeed the faint end slope of the UV luminosity function, sensitive to the dwarf galaxy population, of the Virgo cluster and of the field are very
similar, indicating that other processes such as the formation of dwarfs through merging events (which should induce a flattening of the faint end slope of the Virgo luminosity
function) or the interaction of massive systems (which would produce tidal dwarfs, thus steepening the cluster luminosity function with respect to the field) do not need to be invoked\footnote{
Tidal dwarfs such as those produced in the interacting system Arp 245 (Duc et al. 2000) would be easily detected by GALEX since with FUV and NUV magnitudes of $\sim$ 16-17 mag and central surface brightnesses of 23-24 mag
arcsec$^{-2}$. This is also the case for the star forming blobs in the tail of IC 3418 (Fumagalli et al. 2010), potential progenitors of tidal dwarfs, detected by the GALEX pipeline with UV magnitudes $<$ 21.}.
Given the high velocity dispersion of the cluster, gravitational interactions necessary to produce tidal dwarfs or merging events are extremely rare (Boselli \& Gavazzi
2006). In a ram pressure stripping scenario the decreasing contribution of star forming systems at low luminosities with respect to the field, observed also in the optical bands, 
would be counterbalanced by the increase of the number of quiescent dwarf ellipticals. 
This picture is also consistent with the presence of a significant number of rotationally supported dwarf ellipticals within Virgo
characterized by disky morphologies and young stellar populations (Michielsen et al. 2008; Toloba et al. 2009, 2010). Although the presence 
of dwarf quiescent systems with different morphological properties has been confirmed (Lisker et al. 2006a, 2006b, 2007, 2008, 2009), supporting the idea that different formation processes
at different epochs might have shaped the dwarf galaxy evolution (Lisker 2009), the transformation of low-luminosity star forming systems into dE-dS0 through a ram-pressure 
stripping event is a very plausible scenario for explaining the formation of the low-luminosity extension of the red sequence at recent epochs (Boselli et al. 2008a; Gavazzi et al. 2010). 
The transformation of star forming into quiescent dwarf systems, however, does not seem to work for Coma and A1367, 
where the slope of the UV luminosity function is much steeper than in the field. This observational evidence is inconsistent with the idea that ram pressure stripping is much more efficient in these 
high density clusters than in Virgo (Boselli \& Gavazzi 2006). As previously mentioned, however, statistical corrections necessary at low luminosities combined with a $\sim$ 4 mag brighter cut make 
the fitting parameters of UV luminosity functions of Coma and A1367 still highly uncertain for studying the dwarf galaxy population.

\section{Conclusion}

We have described the GUViCS (GALEX Ultraviolet Virgo Cluster Survey) survey aimed at covering in the UV the whole Virgo cluster 
region ($\sim$ 120 sq. deg.) at the depth of the Medium Imaging Survey (1 orbit per pointing, $\sim$ 1500 sec). 
The data, already available for the central 12 sq. deg., have been used to determine for the first time the FUV and NUV luminosity function of 
galaxies belonging to the core of the Virgo cluster down to the absolute magnitude limit of $M$ $\sim$ -10. This has been done using data for 135 and 65 galaxies in the
NUV and FUV bands respectively. 
Despite the relatively good statistics, the determination of the NUV and FUV luminosity functions is mainly limited by the small sampled volume of the cluster
which drastically limits the number of bright galaxies, making the measure of $M^*$ quite uncertain. Complete UV coverage of Virgo
is necessary for an accurate determination of the UV luminosity function of the cluster.
Our analysis, however, suggests that the FUV and NUV luminosity functions of Virgo galaxies both have slopes $\alpha$ $\sim$ -1.1, significantly 
flatter than that determined for the nearby clusters Coma and A1367
($\alpha$ $\sim$ -1.6) and is rather similar to that measured in the optical B band in Virgo ($\alpha$ = -1.25) or in the UV bands in the field ($\alpha$ $\sim$ -1.2).
Besides the large uncertainty on the determination of $M^*$, the observed difference with Coma and A1367 could be partly attributed to either the much smaller dynamic 
range sampled ($\sim$ 4 mag) and/or the quite uncertain statistical corrections applied for determining the contribution of faint members of these 2 clusters. 
While late-type systems are dominating at bright luminosities, quiescent dE-dS0 are more frequent at low luminosities.\\
The observed shape of the Virgo UV luminosity function, combined with other multifrequency observations and with our spectro-photometric models of galaxy evolution, 
are consistent with the idea that low-luminosity star forming systems recently entered the Virgo cluster and lost
their gas content after the interaction with the hostile environment, quenching their activity of star formation and leading to the formation of quiescent systems.

\begin{acknowledgements}

We wish to thank the GALEX Time Allocation Commetee for the generous allocation of time devoted to this project.
We want to thank J.C. Mu\~noz-Mateos for providing us the SDSS data of the SINGS galaxies and M. Seibert for providing us with FUV data manually reduced
of fields N. 1, 7 and 11 and M. Balogh and the anonymous referee for useful comments. 
This research has made use of the NASA/IPAC Extragalactic Database (NED) 
which is operated by the Jet Propulsion Laboratory, California Institute of 
Technology, under contract with the National Aeronautics and Space Administration
and of the GOLDMine database (http://goldmine.mib.infn.it/).
GALEX (Galaxy Evolution Explorer) is a NASA Small Explorer, launched in 2003 April. We gratefully acknowledge NASA's support for construction, operation, and 
science analysis for the GALEX mission, developed in cooperation with the Centre National d'Etudes Spatiales of France and the Korean Ministry of Science and Technology.
Funding for the SDSS and SDSS-II has been provided by the Alfred P. Sloan Foundation, the Participating Institutions, the National
Science Foundation, the U.S. Department of Energy, the National Aeronautics and Space Administration, the Japanese Monbukagakusho, the
Max Planck Society, and the Higher Education Funding Council for England. The SDSS Web Site is http://www.sdss.org/.
The SDSS is managed by the Astrophysical Research Consortium for the Participating Institutions. The Participating Institutions are the
American Museum of Natural History, Astrophysical Institute Potsdam, University of Basel, University of Cambridge, Case Western Reserve
University, University of Chicago, Drexel University, Fermilab, the Institute for Advanced Study, the Japan Participation Group, Johns
Hopkins University, the Joint Institute for Nuclear Astrophysics, the Kavli Institute for Particle Astrophysics and Cosmology, the Korean
Scientist Group, the Chinese Academy of Sciences (LAMOST), Los Alamos National Laboratory, the Max-Planck-Institute for Astronomy (MPIA), the
Max-Planck-Institute for Astrophysics (MPA), New Mexico State University, Ohio State University, University of Pittsburgh,
University of Portsmouth, Princeton University, the United States Naval Observatory, and the University of Washington.
RG and MPH are supported by NSF grant AST-0607007 and by a grant from the Brinson Foundation.

\end{acknowledgements}

\end{document}